\documentclass{aastex631}
\usepackage{bm}
\usepackage{amsmath}
\usepackage{upgreek}
\usepackage{acronym}
\let\tablenum\relax	
\usepackage[print-unity-mantissa=false]{siunitx}
\usepackage{url}
\usepackage[normalem]{ulem}

\graphicspath{{./}{./}}

\received{January 16, 2022}

\shorttitle{Distributions of \textit{Gaia}-detectable BHBs}
\shortauthors{Shikauchi et al.}

\begin{document}

\title{Spatial and Binary Parameter Distributions of Black Hole Binaries in the Milky Way Detectable with \textit{Gaia}}

\correspondingauthor{Minori Shikauchi}
\email{shikauchi@resceu.s.u-tokyo.ac.jp}

\author{Minori Shikauchi}
\affiliation{Department of Physics, the University of Tokyo, 
7-3-1 Hongo, Bunkyo, Tokyo 113-0033, Japan}
\affiliation{Research Center for the Early Universe (RESCEU), the University of Tokyo,
7-3-1 Hongo, Bunkyo, Tokyo 113-0033, Japan}
\affiliation{Department of Physics and Astronomy, the University of British Columbia, 6224 Agricultural Road, Vancouver, BC, V6T 1Z1, Canada}

\author{Daichi Tsuna}
\affiliation{TAPIR, Mailcode 350-17, California Institute of Technology, Pasadena, CA 91125, USA}
\affiliation{Research Center for the Early Universe (RESCEU), the University of Tokyo,
7-3-1 Hongo, Bunkyo, Tokyo 113-0033, Japan}

\author{Ataru Tanikawa}
\affiliation{Department of Earth Science and Astronomy, College of Arts and Sciences, the University of Tokyo, 
3-8-1 Komaba, Meguro, Tokyo 153-8902, Japan}

\author{Norita Kawanaka}
\affiliation{Center for Gravitational Physics and Quantum Information, Yukawa Institute for Theoretical Physics, Kyoto University, Kitashirakawa Oiwake-cho, Sakyo-ku, Kyoto, 606-8502, Japan}



\begin{abstract}
Soon after the \textit{Gaia} data release (DR) 3 in June 2022, some candidates (and one confirmed) of detached black hole (BH) - luminous companion (LC) binaries have been reported. Existing and future detections of astrometric BH-LC binaries will shed light on the spatial distribution of these systems, which can deepen our understanding of the natal kicks and the underlying formation mechanism of BHs.
By tracking Galactic orbits of BH-LC binaries obtained from \textsf{BSE}, 
we find that distributions of BH mass and the height from the Galactic plane $|z|$ would help us give a constraint on supernova model. 
We also indicate that the correlations of (i) orbital periods and eccentricities, and (ii) BH mass and $|z|$ could be clues for the strength of natal kick, and that the correlations of $(P, Z/Z_\odot)$ may tell us a clue for common envelope (CE) efficiency.
We also discuss the possibility of forming BH-LC binaries like the BH binary candidates reported in \textit{Gaia} DR3 and \textit{Gaia} BH 1, finding that if the candidates as well as the confirmed binary originate from isolated binaries, they favor models which produce low-mass BHs and have high CE efficiencies exceeding unity.
\end{abstract}



\section{Introduction} 
\label{sec:intro}
Massive stars are often formed in binaries, which can leave behind compact objects including black holes (BHs) after core-collapse. Such BHs in binary systems are important tools for probing how BHs are born and evolve, as well as the uncertainties of binary evolution models.
By observing sinusoidal motions of luminous companions (LCs), the astrometric satellite \textit{Gaia} \citep{ESA1997} is supposed to detect non-interacting binaries consisting of LCs and unseen objects, and estimate the mass of the unseen object. If the unseen object mass is larger than a few solar masses and we do not find any excess emission from them by spectroscopy or photometry, the unseen object should be BHs. Since \textit{Gaia} has been observing for more than five years, orbital periods of the detectable binaries with \textit{Gaia} should be tens of days to several years, longer than observed in BH X-ray binaries (XRBs). Observations of low mass XRBs (LMXBs) imply the absence of $3$ -- $5 M_\odot$ BHs \citep{Ozeletal2010,Farretal2011}, so-called lower mass gap \citep{Bailynetal1998}. However, \textit{Gaia} might reveal a completely different BH population from X-ray binaries, and thus has been attracting more and more people's interest. 

There are an increasing number of papers that assess \textit{Gaia}'s detectability of BH-LC binaries \citep[\textit{e.g.}][]{MashianLoeb2017,Breivik2017,Yamaguchi2018,Kinugawa2018,Yalinewichetal2018,Andrewsetal2019,ShaoandLi2019,Wiktorowiczetal2020,Shikauchietal2020,Chawlaetal2021,Shikauchietal2022}. \textit{Gaia} should be able to detect several to thousands of BH-LC binaries in the five-year mission. The number of detectable BH binaries is greatly dependent on some factors such as binary evolution models \citep{Breivik2017,Chawlaetal2021,Shikauchietal2022} and detection criteria adopted in each work.

The recent data release (Data release 3, DR3) was on June 13, 2022 \footnote{https://www.cosmos.esa.int/web/gaia/data-release-3}, which provided about $3.3 \times 10^7$ additional sources from DR2 and the information of $8.1 \times 10^5$ non-single stars, \textit{e.g.}~binaries, from its data spanning about three years.
The \textit{Gaia} collaboration reported BH-main sequence (MS) or post-MS star binary candidates from its spectroscopic data \citep{Arenouetal2022, Gomeletal2022}, which were however rejected by \cite{ElBadryRix2022} to possess BHs for all of the BH-MS star candidates. More recently, \cite{ElBadryetal2023} identified \textit{Gaia} DR3 $4373465352415301632$ (hereafter \textit{Gaia} BH 1) as a binary consisting of a BH and a G dwarf star, and additional BH-LC binary candidates were reported in independent works \citep{Andrewsetal2022, Shahafetal2022, Tanikawaetal2022}. As the number of detections increases in the near future, the distributions of the binary parameters, such as the orbital parameters and locations in the Milky Way (MW) phase-space, would be uncovered. Such distributions should reflect the effect of BH natal kicks that accompany the core-collapse of the BHs' progenitors.

Studies on spatial distributions of BHs have already been done for XRBs \citep{Gandhietal2020, Jonkeretal2021}. 
Analogous to BH XRBs, the spatial distribution of BH binary candidates reported in \textit{Gaia} DR3 may pose an independent constraint on BH natal kick models and their origin, as the \textit{Gaia}-detectable BH-LC binaries are supposed to have longer orbital periods than BH XRBs.

In this work, we investigate the spatial distribution of BH-LC binaries detectable with \textit{Gaia}, by obtaining BH-LC binary population with the binary population synthesis code and tracking their motions under the MW potential. 
In section \ref{sec:method}, we describe the initial spatial condition employed here and the initial set-up for the binary population synthesis code, and explain how to simulate the orbits of BH-LC binaries in the MW from formation to the present day. We show the results in section \ref{sec:result} and compare our samples with the reported BH candidates in section \ref{sec:discussion}. Our conclusion is in section \ref{sec:conclusion}. 

\section{Method} 
\label{sec:method}
In this section, we summarize the initial spatial condition in subsection \ref{ini_conf_mw}. 
The binary population synthesis code and binary evolution models that we employ are depicted in subsection \ref{bps_model}. 
How we track the motion of BH-LC binaries under the MW potential is described in subsection \ref{gal_pot}. We also explain sampling techniques to conduct our simulation efficiently in subsection \ref{sample_tec}.
Finally, the detection criteria with Gaia that are employed in this work are summarized in subsection \ref{subsec:obs_const}.

\subsection{Initial Conditions with Configuration of the MW}
\label{ini_conf_mw}
Here, we follow \cite{Waggetal2021} to synthesize the binary populations throughout the history of the MW.
The formalism of \cite{Waggetal2021} is based on an empirically-informed analytic model that adopts the metallicity-radius-time relations in \cite{Frankeletal2018}. The relations were calibrated based on data of red clump stars observed with APOGEE \citep{Majewskietal2017}.

The MW model consists of three components: the low-[$\alpha$/Fe] disc (\textit{i.e.} the thin disc), the high-[$\alpha$/Fe] disc (\textit{i.e.} the thick disc) and the bar/bulge-like central component. The double disc model reasonably explains the stellar distribution in the MW. 
For the three components, star formation history and the spatial distribution are modelled independently. 
For the star-formation history, we weight each model based on the current stellar mass of each component as follows. 
As of the disc components, the star formation history $p(\tau)$ can be shown as an exponential form,
\begin{eqnarray}
    p(\tau) \mathrm{d}\tau &\propto& \exp \left ( - \frac{\tau_m - \tau}{\tau_\mathrm{SFR}} \right )\mathrm{d}\tau,
\end{eqnarray}
where $\tau$ is the lookback time, \textit{i.e.} the time elapsed from a binary stars' zero-age MS (ZAMS) stage to now, $\tau_m=$12~Gyr is the age of the MW, and $\tau_\mathrm{SFR}$ is a timescale of the star formation, $6.8$~Gyr, based on \cite{Frankeletal2018}.
Note that the periods of star formation in the two discs are different, and stars are formed earlier in the thick disc ($\tau = 8 - 12$~Gyr) and later in the thin disc ($\tau = 0 - 8$~Gyr).
For the bulge component, we adopt a scaled and shifted version of the beta function expressed below following \cite{Waggetal2021}. 
This choice is based on the observations of the Galactic bulge stars. Though there are some uncertainties of the star formation history of the bulge, most of the observed stars in the bulge seem to have ages of $6 - 12$~Gyr. Also, a younger tail of the age distribution comes along, which can be explained by the growth of the bar component \citep[e.g.][]{Bovyetal2019}. Considering above, \cite{Waggetal2021} tried to model the star formation history with the beta function rather than with an old bulge which was created by a single star burst used in previous studies.

In summary, the exact expression of the star formation history $p(\tau)$ including normalization factors is 
\begin{eqnarray}
    p(\tau) \mathrm{d}\tau = \begin{cases}
    \frac{M_\mathrm{disc}}{2 M_\mathrm{tot}} \times n_\mathrm{thin} \exp \left ( - \frac{\tau_m - \tau}{\tau_\mathrm{SFR}} \right ) \mathrm{d}\tau  & (\mathrm{the\;thin\;disc},\;0\;\mathrm{Gyr} \;<\; \tau \;< \;8\;\mathrm{Gyr}), \\
    \frac{M_\mathrm{disc}}{2 M_\mathrm{tot}} \times n_\mathrm{thick} \exp \left ( - \frac{\tau_m - \tau}{\tau_\mathrm{SFR}} \right ) \mathrm{d}\tau  & (\mathrm{the\;thick\;disc},\;8\;\mathrm{Gyr} \;<\; \tau \;< \;12\;\mathrm{Gyr}), \\
    \frac{M_\mathrm{bulge}}{M_\mathrm{tot}} \times \beta(2,3)(\tau') \mathrm{d}\tau & (\mathrm{the\;bulge},\;6\;\mathrm{Gyr} \;<\; \tau \;< \;12\;\mathrm{Gyr}),
\end{cases}
\end{eqnarray}
where the stellar mass of the bulge $M_\mathrm{bulge}$ is $0.9 \times 10^{10} M_\odot$, that of both disc components $M_\mathrm{disc}$ is $5.2 \times 10^{10} M_\odot$ \citep{Licquiaetal2015} assuming the masses of the thin and thick discs are equal \citep[e.g.][]{Snaithetal2014}, and $M_\mathrm{tot} = M_\mathrm{disc} + M_\mathrm{bulge}$, 

\begin{eqnarray}
    \begin{cases}
    n_\mathrm{thin} = \frac{1}{\int^{8\;\mathrm{Gyr}}_{\tau=0\;\mathrm{Gyr}} \exp \left ( - \frac{\tau_m - \tau}{\tau_\mathrm{SFR}} \right ) \mathrm{d}\tau}, \\
    n_\mathrm{thick} = \frac{1}{\int^{12\;\mathrm{Gyr}}_{\tau=8\;\mathrm{Gyr}} \exp \left ( - \frac{\tau_m - \tau}{\tau_\mathrm{SFR}} \right ) \mathrm{d}\tau},
    \end{cases}
\end{eqnarray}
and $\beta(2,3)(\tau')$ is the beta function,
\begin{equation}
    \beta(2,3)(\tau') = \frac{\Gamma(5) \times \tau' (1 - \tau')^2}{\Gamma(2)\Gamma(3)},
\end{equation}
where $\tau' = (\tau/6~\mathrm{Gyr}) - 1$ so that the beta function is scaled and shifted as $\beta=0$ at $\tau=6~\mathrm{Gyr},\;12~\mathrm{Gyr}$ with $\Gamma$, the Gamma function. 

Thus, the number of simulated initial binaries in each component, $N_\mathrm{thin}$ in the thin disc, $N_\mathrm{thick}$ in the thick disc, and $N_\mathrm{bulge}$ in the bulge are
\begin{eqnarray}
\begin{cases}
    N_\mathrm{thin} = N \times \int^{8~\mathrm{Gyr}}_{0~\mathrm{Gyr}} \frac{p(\tau)}{M_\mathrm{tot}} \mathrm{d}\tau = N \times \frac{M_\mathrm{disc}/2}{M_\mathrm{tot}} & (\mathrm{the\;thin\;disc}), \\
    N_\mathrm{thick} = N \times \int^{12~\mathrm{Gyr}}_{8~\mathrm{Gyr}} \frac{p(\tau)}{M_\mathrm{tot}} \mathrm{d}\tau = N \times \frac{M_\mathrm{disc}/2}{M_\mathrm{tot}} & (\mathrm{the\;thick\;disc}), \\
    N_\mathrm{bulge} = N \times \int^{12~\mathrm{Gyr}}_{6~\mathrm{Gyr}} \frac{p(\tau)}{M_\mathrm{tot}} \mathrm{d}\tau = N \times \frac{M_\mathrm{bulge}}{M_\mathrm{tot}} & (\mathrm{the\;bulge}),
    \end{cases}
\end{eqnarray}
where $N=10^7$ is the total number of initial binaries in one realization.

Then, we distribute the initial binaries following the radial and the vertical distributions shown below. 
For the radial distribution, a single exponential distribution is employed, 
\begin{equation}
    q(R) \mathrm{d}R = \exp \left ( - \frac{R}{R_d} \right ) \frac{R}{R_d^2} \mathrm{d}R,
    \label{p_r}
\end{equation}
where $R$ is a radius from the Galactic center, and $R_d$ is a scale length. For the thin disc, $R_d$ is defined as 
\begin{equation}
    R_d \equiv R_\mathrm{exp}(\tau) = 4 \; \mathrm{kpc} \left ( 1 - \alpha_{R_\mathrm{exp}} \left ( \frac{\tau}{8 \; \mathrm{Gyr}} \right ) \right ),
\end{equation}
where $\alpha_{R_\mathrm{exp}} = 0.3$ as the inside-out growth parameter. For the thick disc and the bar structure, $R_d$ is age-independent with the respective values ($1/0.43$)~kpc \citep[Table 1,][]{Bovyetal2019} and $1.5$~kpc \citep{Bovyetal2019}. 

The vertical distribution for each component is a single exponential form as well, 
\begin{equation}
    s(|z|)\mathrm{d}z = \frac{1}{z_d} \exp \left ( - \frac{z}{z_d} \right ) \mathrm{d}z,
    \label{p_z}
\end{equation}
where $z$ is a height from the Galactic plane and $z_d$ is a scale height. The value of $z_d$ for each component is $0.3$~kpc for the thin disc \citep{McMillanetal2011}, $0.95$~kpc for the thick disc \citep{Bovyetal2019}, and $0.2$~kpc for the bulge component \citep{Weggetal2015}.

Finally, the metallicity of each star is given as a function of radius and lookback time, 
\begin{equation}
    [\mathrm{Fe}/\mathrm{H}](R, \tau) = F_m + \nabla [\mathrm{Fe}/\mathrm{H}]  R - \left ( F_m + \nabla [\mathrm{Fe}/\mathrm{H}]  R^{\mathrm{now}}_{[\mathrm{Fe}/\mathrm{H}]=0} \right ) f(\tau),
    \label{metal_rel}
\end{equation}
where 
\begin{equation}
    f(\tau) = \left ( 1 - \frac{\tau}{\tau_m} \right )^{\gamma_{[\mathrm{Fe}/\mathrm{H}]}},
\end{equation}
$F_m=-1$~dex is the metallicity of the star-forming gas at the center of the disc at $\tau = \tau_m$, $\nabla [\mathrm{Fe}/\mathrm{H}]=-0.075$~kpc$^{-1}$ is the metallicity gradient, and $R^{\mathrm{now}}_{[\mathrm{Fe}/\mathrm{H}]=0}=8.7$~kpc is the radius at which the present metallicity is the solar value $Z_\odot = 0.014$.
The value $\gamma_{[\mathrm{Fe}/\mathrm{H}]} = 0.3$ accounts for the time-dependence of the chemical enrichment. 
The metallicity can then be obtained by the relation below \citep[e.g.][]{Bertellietal1994},
\begin{equation}
    \log_{10} (Z) = 0.977 [\mathrm{Fe}/\mathrm{H}] + \log_{10} (Z_\odot). 
\end{equation}
We note that \cite{Waggetal2021} applied this conversion to the thick disc and the bulge component as well as the thin disc, although \cite{Frankeletal2018} fitted this model only for stars in the thin disc. 

To convert the number of BH-LC binaries obtained in the simulation $N_\mathrm{BH-LC, sim}$ to the actual number in the MW $N_\mathrm{BH-LC, MW}$,
\begin{equation}
    N_\mathrm{BH-LC, MW} = N_\mathrm{BH-LC, sim} \times \frac{M_\mathrm{tot}}{M_\mathrm{tot,pop}},
\end{equation}
where $M_\mathrm{tot,pop}$ is the total mass of initial stars we really want to consider in our population synthesis. In the simulation, we prepare initial binaries whose minimum value of the primary ZAMS mass is $8 M_\odot$ while the actual minimum can be as small as $0.08 M_\odot$. Thus, the total mass of initial binaries in the simulation, $M_\mathrm{tot, sim}$ is smaller than that in reality, $M_\mathrm{tot,pop}$. Considering $M_\mathrm{tot, sim} = 3.1 \times 10^8 M_\odot$, the binary fraction is assumed to $0.7$, and the same IMF we employ in the simulation (\textit{i.e.}~Kroupa IMF), the intrinsic total mass in one realization is $1.1 \times 10^9 M_\odot$. As we show in the following section, we duplicate the binary samples by rotating their positions by $2 \pi/50$, we adopt $M_\mathrm{tot,pop}$ as $5.7 \times 10^{10} M_\odot$. The binary fraction is based on the observations of O-type stars in the MW \citep{Sanaetal2012}. 

\subsection{Binary Population Synthesis Code and Binary Evolution Models}
\label{bps_model}
Binary evolution is simulated by the binary population synthesis code \textsf{BSE} \citep{Hurley2000, Hurley2002}. 
We update the stellar wind model in \textsf{BSE} to a metallicity-dependent one following \cite{Belczynski2010}. 

Two different supernova (SN) mechanisms are employed: ``rapid'' and ``delayed'' models suggested in \cite{Fryeretal2012}. 
In the rapid model, BHs as light as $2$ -- $4.5 M_\odot$ are rarely born, which reproduces the lower BH mass gap in X-ray observations \citep{Ozeletal2010,Farretal2011}. Meanwhile, such ``mass gap'' BHs can be formed in the delayed model. 
We use both SN models, as it is still uncertain whether the mass gap is intrinsic or due to observational bias.

We also adopt ``fallback (FB) kick'' model \citep{Fryeretal2012} for the rapid and the delayed SN models as BH natal kicks. The strength of BH natal kicks is that of neutron star (NS) natal kicks modulated by $(1 - f_\mathrm{fb})$, where $f_\mathrm{fb}$ is the fraction of fallback matter to the ejected mass. The distribution of NS kicks is supposed to be Maxwellian distribution with $\sigma=$ \SI{265}{km. s^{-1}} \citep{Hobbsetal2005}. In general, the mass of the remnant BH tends to be larger in the rapid model than in the delayed model, so the magnitude of FB kick in the rapid model is negligible.
In order to see the effect of FB kick, we employ a model with no FB kick for the delayed model as a comparison.
We note that there is also contribution of kick from rapid mass loss upon core-collapse \citep[Blaauw kick;][]{Blaauw1961}, which are included in all of the models.

While the common envelope (CE) phase is treated by $\alpha \lambda$ prescription \citep[equation 3 in][]{Ivanovaetal2013}, two different CE efficiencies, $\alpha = 1$ and $10$ are employed. 
The latter choice is motivated by \cite{ElBadryetal2023}, which indicated that \textit{Gaia} BH 1 cannot be formed with $\alpha = 1$ under the assumption of isolated binary origin, and \cite{HiraiandMandel2022}, which revealed that under their new CE formalism post-CE separations can get as large as those translating to a high CE efficiency reaching $\alpha=10$.
We apply the result in \cite{Claeysetal2014} for $\lambda$. 
In an energy conservation equation in \cite{Webbink1984,Ivanovaetal2013}, $\lambda$ reflects the effect of mass distribution and a contribution from the internal energy of the common envelope \citep{deCool1990, DewiandTauris2000}.

For the distributions of initial binary parameters, we assume a single initial primary mass function of \cite{KroupaIMF2001} from $8 M_\odot$ to $150 M_\odot$. 
The mass ratio is assumed to be flat from $0.1/m_\mathrm{prim, ZAMS}$ to $1$ \citep{Kuiper1935,KobulnickyandFryer2007}. The minimum value of the initial secondary mass is set to $0.1 M_\odot$. 
We also set logarithmically flat distribution for a semi-major axis with a range of $10 R_\odot$ to $10^6 R_\odot$. The initial eccentricity is supposed to be thermally distributed \citep{Heggie1975}. 
As mentioned in subsection \ref{ini_conf_mw}, we track the evolution of $10^7$ initial binaries per each SN/kick model and a choice of $\alpha$. At the beginning of the binary evolution, both stars are in the ZAMS stage. 

\subsection{Tracking the Motion of BH-LC Binaries}
\label{gal_pot}
For those that survive as BH-LC binaries in the present day, we calculate the motion of each binary in the Galaxy from BH formation to today. We follow the formulations of \cite{Tsunaetal2018}, which numerically solved the orbits of isolated BHs under the Galactic potential of \cite{Irrgangetal2013} (their Model II) that contains a spherical bulge, disc and spherical halo. The numerical code calculates the orbit using the cylindrical coordinates $(R,\phi,z)$, with a 4~th-order Runge-Kutta integration. 

\begin{figure}[h]
	\centering
	\plotone{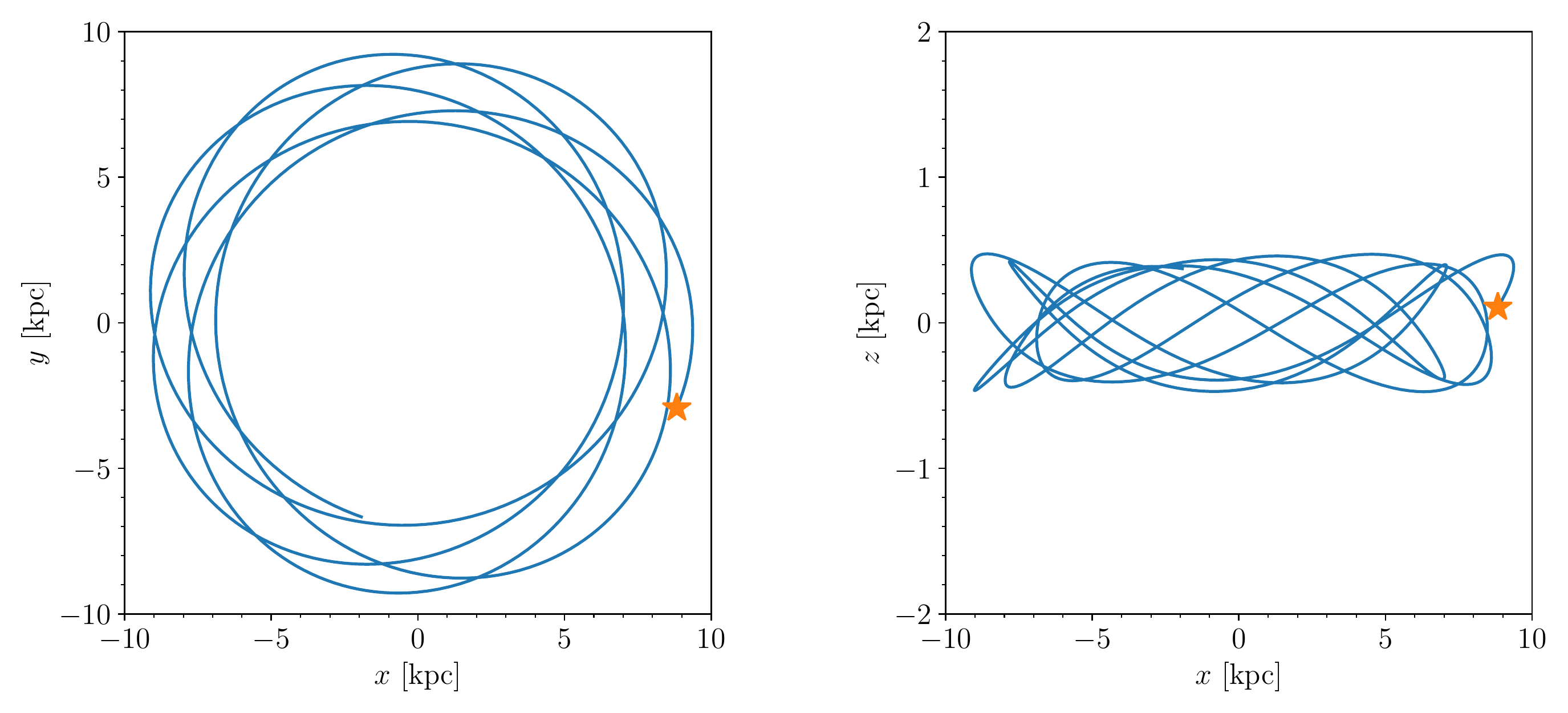}
    \caption{
    An example of the Galactic path of a BH-LC binary in $x-y$ plane (the Galactic plane) and $x-z$ one. The star markers show the initial location of the binary. The path is tracked for $1$~Gyr from now (\textit{i.e.} $\tau = 0$).}  \label{gal_path_circular}
\end{figure}

The displacement of the binary from its birth to BH formation is neglected, and we set the initial $R$ and $z$ coordinates to be those of the binary. Since both the binaries and the Galactic potential follow axisymmetric distributions, we randomize the initial azimuthal angle $\phi$ from $0$ to $2 \pi/50$. That enables us to increase the number of BH-LC binary samples effectively (see section~\ref{sample_tec}). 
We define the initial velocity of the binary by adding the kick to the Galactic rotation velocity, approximated by a rotation curve that is derived from the Galactic potential of \cite{Irrgangetal2013},
\begin{eqnarray}
\upsilon_\phi(r)=
\begin{cases}
	265-1875(r_{\rm kpc}-0.2)^2&\SI{}{km. s^{-1}}\ \ \ ($for$\  r_{\rm kpc}<0.2)\\
	225+15.625(r_{\rm kpc}-1.8)^2&\SI{}{km. s^{-1}}\ \ \ ($for$\  0.2<r_{\rm kpc}<1.8)\\
	225+3.75(r_{\rm kpc}-1.8)&\SI{}{km. s^{-1}}\ \ \ ($for$\  1.8<r_{\rm kpc}<5.8)\\
	240&\SI{}{km. s^{-1}} \ \ \ ($for$\  r_{\rm kpc}>5.8),
\end{cases}
\label{eq:rotation_curve}
\end{eqnarray}
where $r_{\rm kpc}\equiv r/(1\ {\rm kpc})$. For each binary we consider 10 randomized realizations of the kick orientation, assuming it follows an isotropic distribution.
Figure~\ref{gal_path_circular} is an example of the Galactic path of a BH-LC binary in the $x-y$ (\textit{i.e.}~the Galactic plane) and $x-z$ planes. The star marker corresponds to the starting point of the binary at BH formation.

\subsection{Effective Sampling Technique}
\label{sample_tec}
In order to perform our simulation efficiently, we employ two sampling techniques in spatial and temporal ways.
First, we utilize the fact that both the binary distribution and the Galactic potential adopted in our work are axisymmetric. The azimuthal angle distribution of initial binaries are limited to $0$ -- $2 \pi/50$. After tracking the motions of them, we then move the azimuthal angle of the binaries by $2 \pi/50$, and repeat that for $2\pi/(2\pi/50)=50$ times. 
The number of rotations is chosen so that the final number of detectable binaries sufficiently converge.
This sampling technique enables us to increase the number of initial samples to $50 \times 10^7$ with 10 randomized realizations of the kick orientation. 

Our previous work \citep[][]{Shikauchietal2022} found that massive stars with short lifetimes significantly contribute to the luminous sources detectable with \textit{Gaia}, owing to their much larger luminosity. We thus take an importance sampling approach\footnote{\url{https://en.wikipedia.org/wiki/Importance_sampling\#Application_to_simulation}} by employing a bias factor $b(\tau)$,
\begin{eqnarray}
    b(\tau) = \begin{cases}
    N \times f \times n_\mathrm{young} \exp \left ( - \frac{\tau_m - \tau}{\tau_\mathrm{SFR}} \right ) & (\mathrm{the\;thin\;disc},\;0\;\mathrm{Gyr} \;<\; \tau \;< \;0.1\;\mathrm{Gyr}), \\
    N \times (1 - f) \times \frac{M_\mathrm{disc}/2}{M_\mathrm{tot}} \times n_\mathrm{older} \exp \left ( - \frac{\tau_m - \tau}{\tau_\mathrm{SFR}} \right ) & (\mathrm{the\;thin\;disc},\;0.1\;\mathrm{Gyr} \;<\; \tau \;< \;8\;\mathrm{Gyr}), \\
    N \times (1 - f) \times \frac{M_\mathrm{disc}/2}{M_\mathrm{tot}} \times n_\mathrm{thick} \exp \left ( - \frac{\tau_m - \tau}{\tau_\mathrm{SFR}} \right ) & (\mathrm{the\;thick\;disc},\;8\;\mathrm{Gyr} \;<\; \tau \;< \;12\;\mathrm{Gyr}), \\
    N \times (1 - f) \times \frac{M_\mathrm{bulge}}{M_\mathrm{tot}} \times \beta(2,3)(\tau) & (\mathrm{the\;bulge},\;6\;\mathrm{Gyr} \;<\; \tau \;< \;12\;\mathrm{Gyr}).
    \end{cases}
    \label{bias}
\end{eqnarray}
where 
\begin{eqnarray}
    \begin{cases}
    n_\mathrm{young} = \frac{1}{\int^{0.1\;\mathrm{Gyr}}_{\tau=0\;\mathrm{Gyr}} \exp \left ( - \frac{\tau_m - \tau}{\tau_\mathrm{SFR}} \right ) \mathrm{d}\tau}, \\
    n_\mathrm{older} = \frac{1}{\int^{8\;\mathrm{Gyr}}_{\tau=0.1\;\mathrm{Gyr}} \exp \left ( - \frac{\tau_m - \tau}{\tau_\mathrm{SFR}} \right ) \mathrm{d}\tau}, 
    \end{cases}
\end{eqnarray}
and $f$ is a weight factor, here we adopt $0.5$. This biased function $b(\tau)$ shows that 50~\% of the total initial binaries with lookback time restricted to $\tau<0.1$~Gyr, and the rest is assigned to thin disc with $\tau>0.1$~Gyr, thick disc, and the bulge component. 
After the simulation, BH binary samples are obtained and some are considered as detectable based on the detection criteria shown in the following section. 
Considering they are biased following equation~\ref{bias}, we calculate the expected number of detectable BH binaries with the intrinsic star formation history by summing up the ``weighting factor'' $w(\tau)$ for the biased and detectable BH samples,   
\begin{eqnarray}
    w(\tau) &=& \frac{p(\tau)}{b(\tau)} \\
    &=& \begin{cases}
    \frac{\frac{M_\mathrm{disc}}{2M_\mathrm{tot}} n_\mathrm{thin}}{f n_\mathrm{young}} & (\mathrm{the\;thin\;disc},\;0\;\mathrm{Gyr} \;<\; \tau \;< \;0.1\;\mathrm{Gyr}), \\
    \frac{n_\mathrm{thin}}{(1 - f) n_\mathrm{older}} & (\mathrm{the\;thin\;disc},\;0.1\;\mathrm{Gyr} \;<\; \tau \;< \;8\;\mathrm{Gyr}), \\
    \frac{1}{1 - f} & (\mathrm{the\;thick\;disc}), \\
    \frac{1}{1 - f} & (\mathrm{the\;bulge}).
    \end{cases}
    \label{weight}
\end{eqnarray}

\subsection{The Detection Criteria}
\label{subsec:obs_const}
After obtaining the present-day location of BH-LC binaries, we calculate their detectability with \textit{Gaia} by imposing the detection criteria of \cite{Yamaguchi2018} and \cite{Shikauchietal2022}\footnote{Note that \textit{Gaia} BH 1, the confirmed BH-LC binary in \textit{Gaia} DR3 \citep{ElBadryetal2023}, is correctly flagged as detectable by our detection criteria.}. 

We employ three constraints and obtain the maximum distance $D_\mathrm{max}$ within which each BH binary can be detected. If the distance to the BH binary $D$ is smaller than $D_\mathrm{max}$, we regard them as detectable.

\subsubsection{Limitation from Interstellar Extinction}
\label{sec:stellarext}
The first restriction is that the apparent magnitude of a LC $m_{\mathrm{V}}(L_{\mathrm{LC}},T_\mathrm{eff,LC},D_{\mathrm{LC}},z_{\mathrm{LC}})$ should be smaller than \textit{Gaia}'s limiting magnitude in G band $m_\mathrm{v,lim} = 20$ \citep{GaiaColl2016}, that is, 
\begin{equation}
    m_{\mathrm{V}}(L_{\mathrm{LC}},T_\mathrm{eff,LC},D_{\mathrm{LC}},z_{\mathrm{LC}}) = m_{\mathrm{v,lim}},
\end{equation}
where $L_\mathrm{LC}$ is the LC luminosity, $T_\mathrm{eff, LC}$ is the effective temperature of a LC, $D_\mathrm{LC}$ is the maximum distance where the LC satisfies this condition and $z_{\mathrm{LC}}$ is the height of the LC from the Galactic plane. 

The absolute magnitude of a LC $M_\mathrm{V}(L_{\mathrm{LC}},T_\mathrm{eff,LC})$ can be obtained from $L_{\mathrm{LC}}$ and $T_\mathrm{eff,LC}$ with a bolometric correction \citep[\textit{c.f.} equation 1, 10, and Table 1 in][]{Torres2010}. Note that we substitute G band with V band. This is a valid approximation for stars bluer than G type stars whose color $V-I$ is less than one and the color $|V - G|$ is almost zero according to Figure 11 and 14 of \cite{Jordietal2010}. 
The apparent magnitude of a LC $m_{\mathrm{V}}$ is expressed as a function of the distance to BH binary $D$ and the height from the Galactic plane to the binary $z$, 
\begin{equation}
	m_\mathrm{v} = M_\mathrm{V}(L_{\mathrm{LC}},T_\mathrm{eff,LC}) + 5(2 + \log_{10}D/\mathrm{kpc}) + A_\mathrm{V}(D,z),
\label{ext}
\end{equation}
where $D/\mathrm{kpc}$ is $D$ in units of kpc. 
The term $A_\mathrm{V}$ due to interstellar extinction can be expressed following \cite{Shafter2017},
\begin{eqnarray}
    A_\mathrm{V}(D,z) &=& a_\mathrm{V} \int^{D}_0 \mathrm{e}^{-|z|/h_z} \mathrm{d}D' \\
    &=& a_\mathrm{V} \frac{D h_z}{|z|} \left [ 1 - \exp \left ( -\frac{|z|}{h_z} \right ) \right ],
\end{eqnarray}
where $a_\mathrm{V}$ is the average extinction rate in the Galactic plane ($z=0$), $1$ mag/kpc, and $h_z=100$ pc is the scale height in the $z$-direction perpendicular to the plane \citep{Spitzer1978}.
Thus, the maximum distance satisfying the condition $D_{\mathrm{LC}}$ is
\begin{equation}
	M_{\rm V}(L_{\mathrm{LC}},T_\mathrm{eff,LC}) + 5(2 + \log_{10}D_{\rm LC}/\mathrm{kpc}) + A_\mathrm{V}(D_{\rm LC}, z_\mathrm{LC}) = m_{\mathrm{V, lim}}.
 \label{int_ext}
\end{equation}
Note that $D_\mathrm{LC}$ depends on the line-of-sight angle with respect to the plane, since the extinction term $A_\mathrm{V}$ depends on $z_\mathrm{LC}$.

\subsubsection{Constraints for Confirmed Detection of BHs}
\label{sec:obsconst}
In astrometric observations, we can only identify BHs or NSs based on their masses. In order to consider unseen objects as BHs, we restrict the minimum mass of them to be measured as larger than $2 M_\odot$,   
\begin{equation}
    m_{\mathrm{unseen}} - n \sigma_{\mathrm{unseen}} > 2M_{\odot},
\label{BHconst}
\end{equation}
where $m_{\mathrm{unseen}}$ is their true mass and $\sigma_{\mathrm{unseen}}$ is its standard error. We follow \cite{Yamaguchi2018} and adopt $n=1$.
Though the minimum limit we set here may induce contamination of NSs, searching for compact objects with masses of $2 - 3 M_\odot$ should be valuable as the existence of such an object was reported in gravitational wave searches \citep[GW190814, ][]{Abbottetal2020GW190814}. 

From Kepler's third law the binary parameters, LC mass $m_\mathrm{LC}$, BH mass $m_\mathrm{BH}$, orbital period $P$ and semi-major axis $a$, are correlated. Considering that $a$ can be expressed by a multiplication of an angular semi-major axis $a^*$ and the distance to BH-LC binary $D$, the correlation of binary parameters is shown as 
\begin{equation}
    {\frac{(m_{\mathrm{LC}} + m_{\mathrm{BH}})^2}{m_{\mathrm{BH}}^3}} = {\frac{G}{4 \pi^2}} {\frac{P^2}{(a_* D)^3}},
\label{binary}
\end{equation}
where $G$ is the gravitational constant. 
Ignoring the correlation of each parameter and observational errors, we derive a relationship between each parameter and its standard error, 
\begin{equation}
     \left( {{\sigma_{\mathrm{BH}}}\over{m_{\mathrm{BH}}}} \right )^2 = \left( {{3}\over{2}} - {{m_{\mathrm{BH}}}\over{m_{\mathrm{BH}}+m_{\mathrm{LC}}}} \right )^{-2} \left[ \left( {{m_{\mathrm{LC}}}\over{m_{\mathrm{BH}}}+m_{\mathrm{LC}}} \right )^2 {{\sigma_{\mathrm{LC}}^2}\over{m_{\mathrm{LC}}^2}} + {{\sigma_{P}^2}\over{P^2}} + {{9}\over{4}} \left( {{\sigma_{a*}^2}\over{a_*^2}} + {{\sigma_{D}^2}\over{D^2}} \right) \right].
     \label{mbh}
\end{equation}
where $\sigma$ is a standard error and each suffix corresponds to each binary parameter. 

For confident detection of BHs, we impose a condition that the error of each parameter must be smaller than $10$~\% of the true value, 
\begin{equation}
    {{\sigma_{\mathrm{LC}}}\over{m_{\mathrm{LC}}}} < 0.1, 
    {{\sigma_{P}}\over{P}} < 0.1, {{\sigma_{a*}}\over{a_*}} < 0.1,
    \;\; \mathrm{and } \; \; {{\sigma_{D}}\over{D}} < 0.1.
    \label{const0}
\end{equation}
Under these requirements, detection of BHs with $m_{\rm BH}\gtrsim 3.4 M_\odot$ should be confirmed as BHs. 

The conditions for LC mass and orbital period are easily satisfied. According to \cite{Tetzlaffetal2011}, a standard error of LC mass based on its spectrum and luminosity is typically smaller than 10~\%. Furthermore, the standard error of orbital periods is suppressed to below 10~\% if the observed periods are shorter than $2/3$ of the operation time of \textit{Gaia} \citep{ESA1997}. As \cite{Lucy2014} and \cite{ONeiletal2019} proposed a novel technique to estimate binary parameters when the orbital coverage is less than 40~\%, and \textit{Gaia} has been observing for more than five years, we employ $10$~years as the maximum period of observable BH-LC binaries. For the lower limit of orbital periods, we set 50~days as \cite{Yamaguchi2018} does. 
The rest of the conditions in equation~(\ref{const0}) impose two more constraints on $D_\mathrm{max}$. 
First, considering that the parallax $\Pi$ is proportional to the reciprocal of $D$, the ratio of the standard error of parallax $\sigma_\Pi$ and $\Pi$ can be approximated to that of $\sigma_D$ and $D$, 
\begin{equation}
  \frac{\sigma_\Pi}{\Pi} \sim \frac{\sigma_D}{D} < 0.1.
  \label{eq:ParallaxAndDistance}
\end{equation}
\cite{GaiaColl2016} provided $\sigma_\Pi$ in G band as a function of the apparent magnitude of a LC $m_{\mathrm{v}}$ and we employ the expression below ignoring the dependence on the color $V-I$,  
\begin{equation}
    \sigma_\Pi = (-1.631 + 680.8z(m_\mathrm{v}) +
    32.73z(m_{\mathrm{v}})^2)^{1/2} [\mu \mathrm{as}], 
    \label{eq:StandardErrorOfParallax}
\end{equation}
where
\begin{equation}
    z(m_{\mathrm{v}}) = 10^{0.4 (\mathrm{max}[12.09,m_{\mathrm{v}}]-15)}.
\end{equation}
Combining equations~(\ref{eq:ParallaxAndDistance}) and (\ref{eq:StandardErrorOfParallax}), the second constraint for $D_\mathrm{max}$ is 
\begin{equation}
	\left( \frac{D_\mathrm{max}}{\rm kpc} \right) < D_{\Pi} = \frac{10^2}{(-1.631 + 680.8z(m_\mathrm{v}) + 32.73z(m_{\mathrm{v}})^2)^{1/2}}.
  \label{parallax_err}
\end{equation}

Finally, for the condition of angular semi-major axis, we approximate the uncertainty of the angular semi-major axis of a BH binary $\sigma_{a*}$ as that of its orbital radius on the celestial sphere $\sigma_{\Pi}$. Then, the final condition for $D_\mathrm{max}$ can be obtained,  
\begin{equation}
        \left( \frac{D_\mathrm{max}}{\rm kpc} \right) < D_{a} = \frac{am_{\rm BH}}{10(m_{\rm BH}+m_{\rm LC}) \sigma_\Pi}.
    \label{sep_err}
\end{equation}

In summary, we obtain three constraints for $D_\mathrm{max}$, $D_\mathrm{LC}$ (equation~\ref{int_ext}) $D_\Pi$ (equation~\ref{parallax_err}), and $D_{a}$ (equation~\ref{sep_err}). For each BH binary sample, we compare the minimum of the three to the current distance to determine whether the binary is detectable. 

We call the above detection criteria as ``optimistic'', since the errors for the parameters are determined independently.
However, in realty, this would not be the case since all the binary parameters come from observations of LCs and covariance terms should be included in equation~(\ref{mbh}). 
In order to consider the covariance terms that we ignored in the derivation, we consider a modified detection criteria based on the observations of the confirmed BH binary \textit{Gaia} BH 1 \citep{ElBadryetal2023}.
For \textit{Gaia} BH 1, $\sigma_\mathrm{BH}$ in \cite{ElBadryetal2023} is about three times larger than estimated in equation~(\ref{mbh}), which means we adopt three times more optimistic detection criteria.
The value $\sigma_\Pi$ is 12 times smaller and $\sigma_{a*} (\sim \sigma_\Pi)$ is $\sim 1.36$ times larger than estimated in equation~(\ref{eq:StandardErrorOfParallax}).
In order to obtain a close value of $\sigma_{\mathrm{BH}}$ and retain that \textit{Gaia} BH 1 is considered as detectable under the criteria, 
we employ the detection criteria by adopting $\sigma_\Pi/\Pi < 0.025$ and $\sigma_{a*}/a_* < 0.1/1.36$, that is,
\begin{equation}
    {{\sigma_{\mathrm{LC}}}\over{m_{\mathrm{LC}}}} < 0.1, 
    {{\sigma_{P}}\over{P}} < 0.1, {{\sigma_{a*}}\over{a_*}} < 0.1/1.36,
    \;\; \mathrm{and } \; \; {{\sigma_{\Pi}}\over{\Pi}} < 0.025.
    \label{const}
\end{equation}
which we call ``conservative'' detection criteria for comparison. 

\section{Result} 
\label{sec:result}
Based on the results of \textsf{BSE} and the orbit calculations, we obtain the spatial distributions and binary parameters of the Galactic BH-LC binaries. 
We summarize in Table \ref{det_num} the number of BH-LC binaries in the MW with orbital periods of $50$~days to $10$~years, $N_\mathrm{BH-LC,MW}$, and the number of detectable BH binaries, $N_\mathrm{det}$, for each SN/kick model and a choice of $\alpha$. The number of detectable binaries for each model is several times larger than estimated in our previous work \citep{Shikauchietal2022}, which can be explained by the following differences between the two works. 
In this work, we have considered a realistic star formation history instead of a constant star formation rate. That drastically increases the number of BH binaries with low mass LCs ($m_\mathrm{LC} \lesssim 1 M_\odot$), and also shows different BH/LC mass distributions from our previous work. Binary and spatial parameter distributions are shown in Appendix~\ref{app:corr}.
Furthermore, while the previous work employed a single metallicity value of solar for all binaries, here we have considered the metallicity to vary as a function of radius and lookback time. As for binaries born in the past with generally lower metallicity, progenitors with smaller ZAMS masses can evolve into BHs instead of NSs due to reduced mass loss. In addition, the number of heavier BHs will increase, which would make the binary easier to detect. 

\begin{table}[h]
	\centering
\begin{tabular}{ccc|cccc} \hline
SN model & kick & $\alpha$ & $N_\mathrm{BH-LC,MW}$ & \multicolumn{2}{c}{$N_\mathrm{det}$} & \cite{Shikauchietal2022} \\ 
  &   &   &   & optimistic & conservative &  \\ \hline
delayed & FB kick & $1$ & $3.15 \times 10^3$ & $7.16^{+3.01}_{-4.44}$  & $4.04^{+2.50}_{-3.59}$ & $1.1$ \\ 
\ldots & no kick & \ldots & $8.59 \times 10^3$ & $40.3^{+4.51}_{-2.01}$ & $22.5^{+3.61}_{-3.93}$ & $22$\\ 
rapid & FB kick & \ldots & $7.53 \times 10^3$ & $44.3^{+1.44}_{-3.22}$ & $16.8^{+2.85}_{-1.65}$ & $18$\\ \hline 
delayed & FB kick & $10$ & $6.33 \times 10^3$ & $10.9^{+3.85}_{-1.03}$ & $4.15^{+4.29}_{-3.01}$ & $9.4$\\ 
\ldots & no kick & \ldots & $2.91 \times 10^4$ & $56.4^{+5.34}_{-2.62}$ & $29.5^{+3.53}_{-1.54}$ & $46$ \\ 
rapid & FB kick & \ldots & $1.18 \times 10^4$ & $67.6^{+4.31}_{-1.71}$ & $31.2^{+1.77}_{-2.94}$ & $31$ \\ \hline 
\end{tabular}
	\caption{The number of BH-LC binaries in the MW $N_\mathrm{BH-LC,MW}$ with $P$ between $50$~days and $10$~years, and those detectable with \textit{Gaia} $N_\mathrm{det}$, for different choices of SN/kick models and values of the CE efficiency $\alpha$ under different choices of the detection criteria. The numbers and errors in $N_{\rm det}$ correspond to the median and the spread between $10$~th and $90$~th percentiles for the 10 realizations of the kick orientation.}
 \label{det_num}
\end{table}

In order to evaluate the expected correlation between each binary parameter and spatial parameters of the intrinsic BH binaries, we employ the ``weighted'' Pearson correlation coefficients,
\begin{equation}
    \rho_{XY, w} = \frac{\mathrm{cov}(X,Y,w)}{\sqrt{\sigma_{X,w} \sigma_{Y,w}}},
\end{equation}
where $X,Y$ are choices of binary parameters and spatial information of the biased BH binary samples, cov$(X,Y,w)$ is a weighted covariance matrix of $X$ and $Y$, 
\begin{equation}
    \mathrm{cov}(X,Y,w) = \frac{\sum\limits_i (w_i \times (X_i - \bar{X}) \times (Y_i - \bar{Y}))}{\sum\limits_i w_i},
\end{equation}
$w$ is the weighting factor for each binary (see equation~\ref{weight}), $\bar{X}, \bar{Y}$ are weighted means of $X$ and $Y$, and $\sigma_{X,w}, \sigma_{Y,w}$ are weighted standard deviations of $X, Y$, \textit{i.e.}~cov$(X,X,w)$ and cov$(Y,Y,w)$. 

The coefficients of the detectable BH-LC binaries with each SN/kick model and value of the CE efficiency $\alpha$ are summarized in the left panels of Figure~\ref{corr_defba1} - \ref{corr_denoa10}. 
Values of the coefficients are categorized to seven levels: ``strongly positive correlation'' ($1.0 \sim 0.7$), ``positive correlation'' ($0.7 \sim 0.4$), ``weakly positive correlation'' ($0.4 \sim 0.2$), ``no correlation'' ($0.2 \sim -0.2$), ``weakly negative correlation'' ($-0.2 \sim -0.4$), ``negative correlation'' ($-0.4 \sim -0.7$), and ``strongly negative correlation'' ($-0.7 \sim -1$).
The right panels of Figure~\ref{corr_defba1} - \ref{corr_denoa10} show correlation coefficients for the entire Galactic binary population with orbital periods from $50$~days to $10$~years, for each SN/kick model and $\alpha$. Most of them show no correlations. 
Correlation coefficients seen in the detectable BH-LC binaries have the opposite sign and/or are enhanced compared with the correlations among the Galactic BH-LC population. Thus, most of the correlations are generally biased by the detection criteria. 

In the following subsections, we look into significant correlations of the detectable BH-LC binaries in each model. 
Note that all the figures~\ref{corr_defba1} - \ref{corr_denoa10} are based on the conservative detection criteria, but the trend explained below are generally retained for the optimistic criteria, albeit slight differences for the correlation coefficients.

One may be concerned about the rotational procedure causing artificial correlations. 
Though the samples are not independent any longer, this will not have a significant effect for the Galactic BH binaries since all the binaries are equally duplicated through the procedure. The correlations should converge to the true ones without the procedure.

In addition, the information we need is the expected number of BH binaries in a given point of the parameter space,
not of the individual binary.
The 50-time rotations are chosen so that the number of BH binaries in the parameter space will converge, and at the same time to avoid producing artificial spatial correlations. 

There are two possibilities which can produce unexpected correlations: a large-scale pattern appearing if we do not rotate the samples very much and a small-scale one if we rotate them too much and oversample the binaries. 
We can estimate the expected scale of the large-scale pattern by calculating the average separation between a given pair of initial binaries. It is $4$--$5$~kpc from the simulation.
On the other hand, the latter constraint for the rotation can be estimated by requiring the distance between adjacent rotations is larger than the average separation of the binaries. Distributing $10^4$ binaries (\textit{i.e.}~the total number of surviving BH binaries in the MW obtained from the simulation) in the disk with a radius of $10$~kpc, the average separation is estimated to be about $0.1$~kpc.
The average distance between the original and the adjacent rotated samples can be expressed as $r \sqrt{2 - 2 \cos(2 \pi / n)}$ where $r$ is the average distance from the Galactic center to the original samples in the Galactic plane and $n$ is the number of rotations. Considering the average $r$ is roughly $5$~kpc, the average distance between adjacent rotations is $\sim 0.6~{\rm kpc}$, which is comfortably between the scale of the smaller structure and the scale of the larger structure with $n = 50$, which means this rotation procedure would not cause any significant artificial correlations. 

However, artificial correlations might arise for the detectable binaries if $N_\mathrm{det}$ is small, due to Poisson fluctuations and/or the choice of detection criteria. 
This can be significant for the models with delayed SN model and FB kick, where we expect to have only a handful of astrometric detections at most.
In order to see the effect of Poisson fluctuation to the coefficients, we checked how the values of the coefficients would fluctuate by calculating the coefficients for each of the 10 different kick realizations.
Picking up 10~\% and 90~\% percentiles of the coefficients among the realizations, we found that strong correlations (i.e. those with absolute values $>0.4$) are preserved in 10~\% and 90~\% percentiles as well. 
Their coefficients may change by $\pm \sim 6$~\% with $\alpha=1$ case, by $\pm \sim 20$~\% with $\alpha=10$ case from the values with all the samples included at the same time. Thus, while these strong correlations are robust against both Poisson fluctuations and choice of the detection criteria, one should be more cautious of our predictions for the weaker correlations when $N_{\rm det}$ is small.
As radial velocity searches can reduce the error of the astrometric mass function by several times \citep[e.g.][]{ElBadryetal2023}, we expect that more BH binaries will be robustly detected than our estimation under the conservative criteria by pure astrometry, and in those cases the weaker correlations may also become statistically significant.

\subsection{the delayed SN model/FB kick/$\alpha=1$}
\label{subsec:delay_FB_alpha1}
\begin{figure}[h]
	\centering
	\plotone{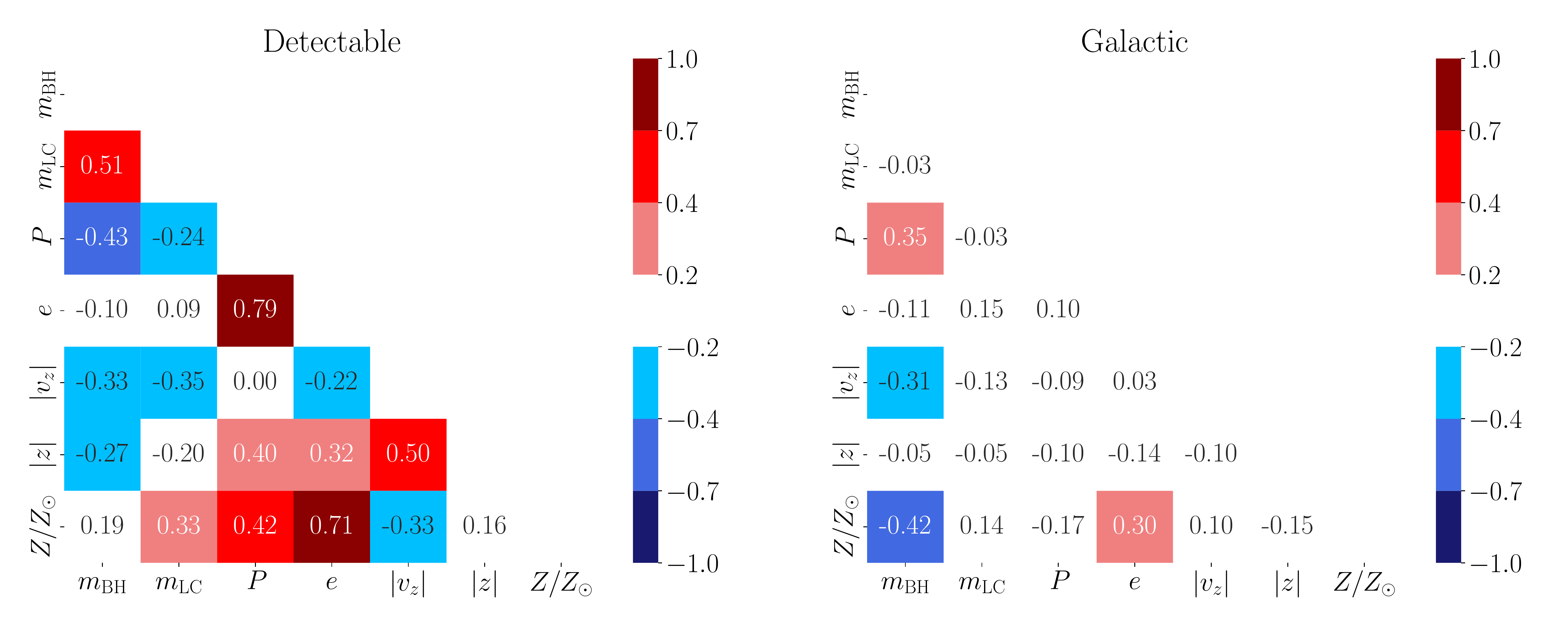}
    \caption{(Left) Correlation coefficients between the current binary parameters (BH mass $m_\mathrm{BH}$, LC mass $m_\mathrm{LC}$, orbital periods $P$, and eccentricities $e$), the current spatial parameters (velocities perpendicular to the Galactic plane $|v_z|$ and the heights from the Galactic plane $|z|$), and metallicity $Z/Z_\odot$ of the detectable BH-LC binaries with the delayed SN model/FB kick/$\alpha=1$ under the conservative detection criteria. (Right) Same as the left panel, but of the entire Galactic BH-LC binary population with orbital periods from $50$~days to $10$~years.}  \label{corr_defba1}
\end{figure}

In the delayed SN model with FB kick and $\alpha=1$, 
\begin{enumerate}
    \item strongly positive correlations of $(P, e)$ and $(e, Z/Z_\odot)$,
    \item positive correlations of $(m_\mathrm{BH}, m_\mathrm{LC})$, $(P, Z/Z_\odot)$, and , $(|z|, |v_z|)$
    \item negative correlations of $(m_\mathrm{BH}, P)$
\end{enumerate}
are seen in Figure~\ref{corr_defba1}. 

The strongly positive correlation of $(P,e)$ can be understood based on the correlations of $(P, Z/Z_\odot), (e, Z/Z_\odot)$. Heavier BH binaries are formed in lower metallicity, suffering from smaller fallback kick. This results in less eccentric and narrower orbits compared to binaries with lighter BHs. 
The detection criteria favor long period binaries, which emphasizes the positive correlation of $(P, Z/Z_\odot)$. The positive correlation of $(m_\mathrm{BH}, m_\mathrm{LC})$ is highlighted by the detection criteria. Heavier BHs can swing around heavier LCs largely and are more detectable.

Although not significant, there are very weak negative correlations of $(m_\mathrm{BH}, |z|)$ and $(m_\mathrm{BH}, |v_z|)$. That can be explained considering lighter BHs suffer from larger natal kicks, leading to move farther away from the Galactic plane. 
That would also explain the correlation of $(|z|, |v_z|)$.
In addition, as seen in the negative correlation of $(m_\mathrm{BH}, P)$, binaries with lighter BHs would have larger orbital separations due to larger kicks. 

\subsection{the rapid SN model/FB kick/$\alpha=1$}
\begin{figure}[h]
	\centering
	\plotone{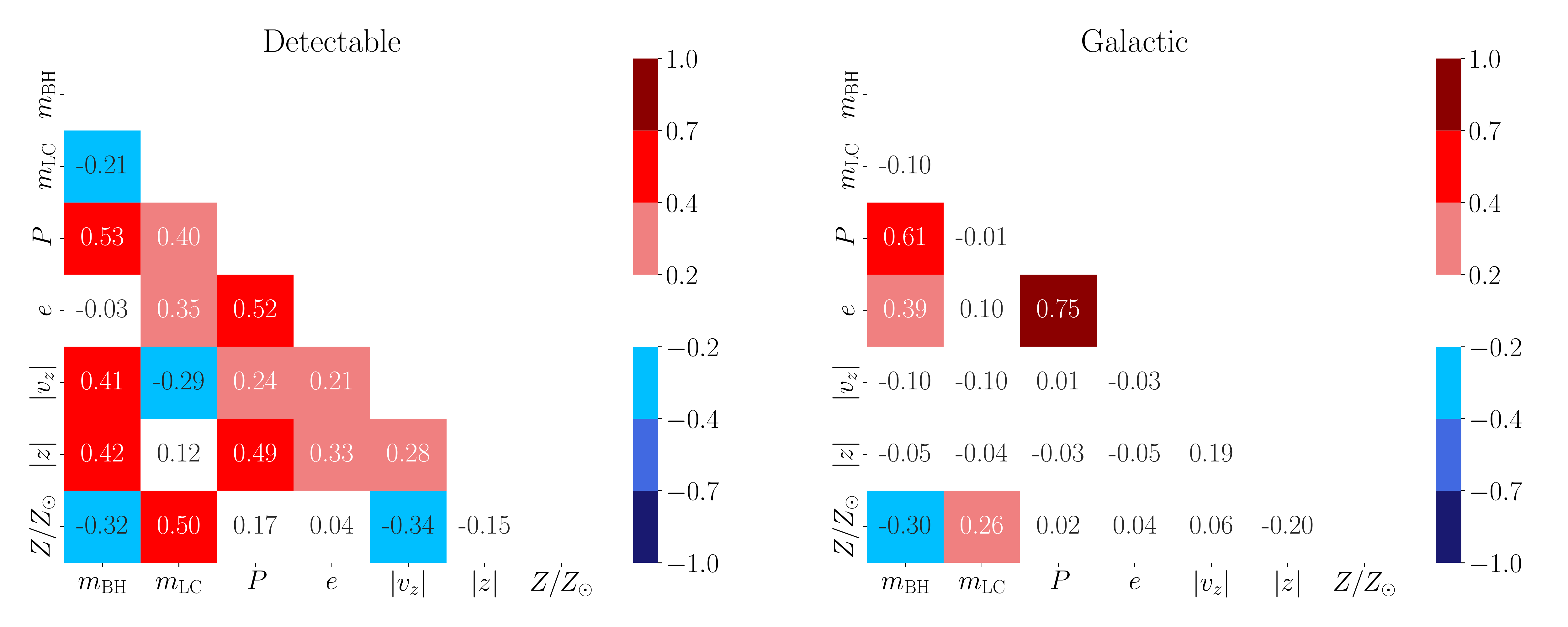}
    \caption{Same as Figure~\ref{corr_defba1}, but with a different SN model, the rapid SN model.}  \label{corr_rafba1}
\end{figure}

In the rapid SN model with FB kick and $\alpha=1$, 
\begin{enumerate}
    \item positive correlations of $(m_\mathrm{BH}, P)$, $(m_\mathrm{BH}, |v_z|)$, $(m_\mathrm{BH}, |z|)$, $(m_\mathrm{LC}, Z/Z_\odot)$, $(P, e)$, and $(P, |z|)$, 
\end{enumerate}
are seen in Figure~\ref{corr_rafba1}. 

The positive correlation of $(m_\mathrm{BH}, P)$ can be interpreted as follows. 
For light BH binaries, BHs are formed after the CE phase. 
On the other hand, heavier BH binaries ($m_\mathrm{BH} \gtrsim 10 M_\odot$) do not experience the CE phase because they cannot survive if they enter the phase as shown below. 
Since heavy BHs are formed in low metallicity, heavy BH binaries are typically born in the distant past. 
They tend to have low mass LCs ($m_\mathrm{LC} \lesssim 1 M_\odot$)
\footnote{The reason we do not see any correlations of $(m_\mathrm{BH}, m_\mathrm{LC})$ among the detectable BH binaries is because not only heavy BH progenitors but also light ones mostly have low-mass LCs. In our MW model, a larger number of binaries with low-mass LCs exist until now compared to those with high-mass LCs. While binaries with low-mass LCs are more difficult to observe than those with high-mass LCs, they outnumber enough to be still dominant among the detectable binaries.}, otherwise they cannot exist as BH-LC binaries until the present day. 
However, the ZAMS masses of the progenitors of these heavy BHs are as large as $\gtrsim$ tens of $M_\odot$.
Here, we roughly estimate the final orbital separations if such high mass ratio binaries enter the CE phase. 
Considering $\alpha \lambda$ prescription of the CE phase, the binding energy of a binary at the beginning of the CE phase is roughly proportional to the orbital energy of a binary at the end of the phase. 
Orbital separations at the final stage of the CE phase $a_\mathrm{f}$ can be approximated as 
\begin{align}
    a_\mathrm{f} &= \frac{m_\mathrm{prim, core} m_\mathrm{second, ZAMS}}{2} \times \left ( \frac{m_\mathrm{prim, i} m_\mathrm{prim, env}}{\alpha \lambda R} + \frac{m_\mathrm{prim, i} m_\mathrm{second, ZAMS}}{2 a_\mathrm{i}} \right )^{-1} \\
    &\sim \frac{\alpha \lambda}{2} \times \frac{m_\mathrm{second, ZAMS} m_\mathrm{prim, core}}{m_\mathrm{prim, i} m_\mathrm{prim, env}} R, 
\end{align}
where we have defined the initial secondary mass $m_\mathrm{second, ZAMS}$, the primary mass at the beginning of the CE phase $m_\mathrm{prim, i}$, the envelope mass of the primary $m_\mathrm{prim, env}$, the core mass of the primary $m_\mathrm{prim, core}$, the orbital separation at the beginning of the phase $a_\mathrm{i}$ and Roche lobe radius of the primary $R$. 
Assuming that mass loss is negligible in low metallicity, $m_\mathrm{prim, i} \sim m_\mathrm{prim, ZAMS}$ and $m_\mathrm{prim, core}/m_\mathrm{prim, env}\sim 0.5$ is almost independent of the primary mass \citep[\textit{e.g.}~section 4.2 in][]{Sukhboldetal2018}. 
Since $R$ is approximated to tens of solar radii and $\lambda \sim 0.4$, $a_\mathrm{f} \lesssim 0.1 R_\odot$ with $\alpha = 1$, and $\lesssim R_\odot$ even for $\alpha = 10$. It is smaller than the core radius of the primary, $\sim R_\odot$, which leads high mass ratio binaries with $m_\mathrm{second, ZAMS}/m_\mathrm{prim, ZAMS}\ll 1$ to merge.
Thus, existing binaries with heavy BHs are limited to have longer orbital periods that do not experience the CE phase. 

The positive correlation of $(P, e)$ exists as well, but can be interpreted in a different way from in the delayed SN model with FB kick. In the rapid SN model, natal kick is not as strong as in the delayed SN model. Thus, BH binaries experiencing the CE phase simply have smaller eccentricities and narrower orbits. 

The weaker natal kick in the rapid model also explains the positive correlation of $(m_\mathrm{BH}, |z|)$. In the delayed SN model with FB kick, lighter BH binaries can move farther away from the Galactic plane due to strong FB kick. However, such light BHs are rarely formed in the rapid SN model and BH binaries do not go farther. Rather, the detection criteria highlight the fact that heavier BH binaries are detectable at farther distances according to equation~\ref{sep_err}. The correlation of $(P, |z|)$ is also highlighted by the detection criteria; binaries with larger orbits can be easily detected according to equation~(\ref{sep_err}). 

The correlation of $(m_\mathrm{BH}, |v_z|)$ can be explained by the correlation of $(m_\mathrm{BH}, |z|)$ and a weak correlation of $(|z|, |v_z|)$, which implies binaries located farther away from the Galactic plane have a large $|v_z|$.

The positive correlation of $(m_\mathrm{LC}, Z/Z_\odot)$ is easily explained by the fact that only massive LCs that were born recently can survive until today.

Comparing with the result in the delayed SN model with FB kick, the correlation coefficients of $(m_\mathrm{BH}, |z|)$ ($-0.27$ with the delayed SN model, $0.42$ with the rapid SN model), and $(m_\mathrm{BH}, |v_z|)$ ($-0.33$ with the delayed SN model, $0.41$ with the rapid SN model), have the opposite signs. 
As mass gap BHs ($m_\mathrm{BH} \lesssim 5 M_\odot$) will be detectable only in the delayed SN model, the distribution of $(m_\mathrm{BH}, |z|)$ and $(m_\mathrm{BH}, |v_z|)$ would be a powerful tool to constrain the SN model. 

\subsection{the delayed SN model/no kick/$\alpha=1$}
\begin{figure}[h]
	\centering
	\plotone{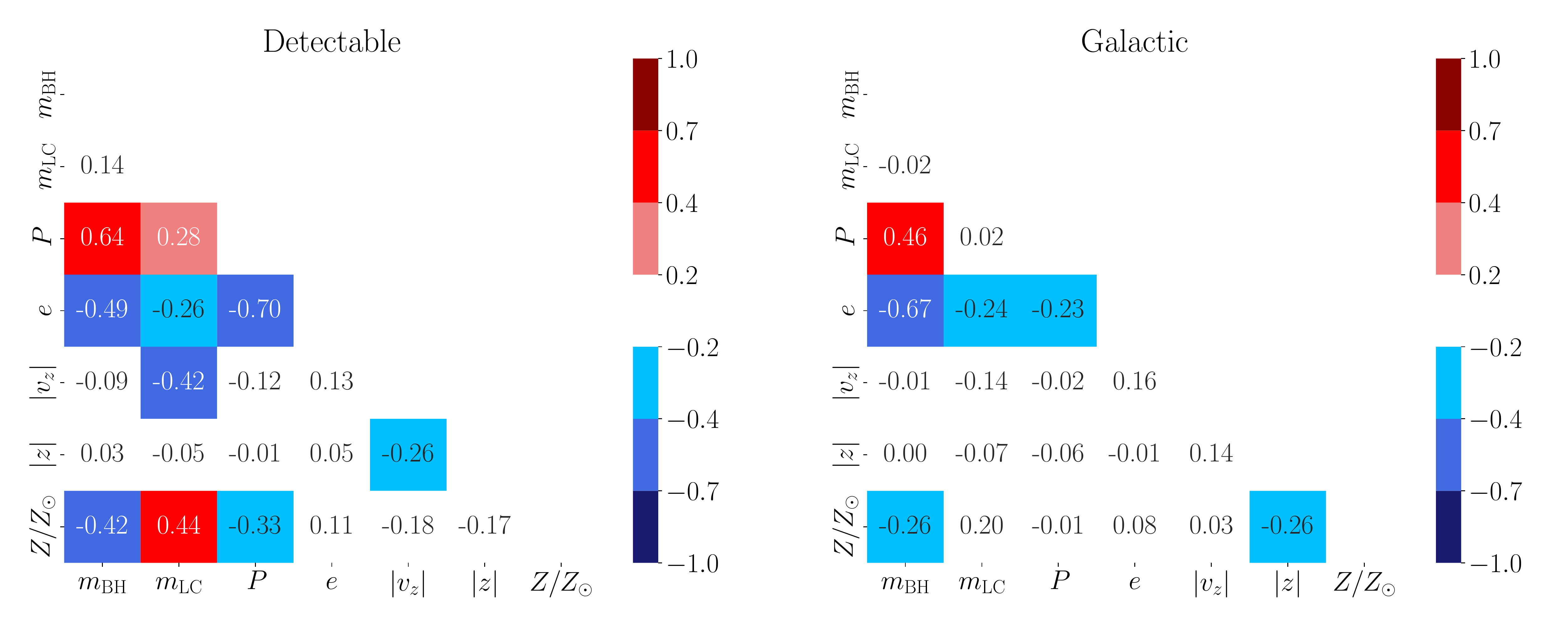}
    \caption{Same as Figure~\ref{corr_defba1}, but with a different kick model, no kick.}  \label{corr_denoa1}
\end{figure}

In the delayed SN model with no kick and $\alpha=1$, there are 
\begin{enumerate}
    \item positive correlations of $(m_\mathrm{BH}, P)$ and $(m_\mathrm{LC}, Z/Z_\odot)$,
    \item negative correlations of $(m_\mathrm{BH}, e)$, $(m_\mathrm{BH}, Z/Z_\odot)$, $(P,e)$, and $(m_\mathrm{LC}, |v_z|)$
\end{enumerate}
in Figure~\ref{corr_denoa1}.

The correlations of $(m_\mathrm{BH}, P)$ and $(m_\mathrm{LC}, Z/Z_\odot)$ show a similar trend seen in the rapid SN model.

The negative correlations of $(m_\mathrm{BH}, e)$ is understandable as lighter BH binaries suffer from larger Blaauw kicks. Combining correlations of $(m_\mathrm{BH}, P)$ and $(m_\mathrm{BH}, e)$, the negative correlation of $(P,e)$ would be reasonable. 

The negative correlation of $(m_\mathrm{LC}, |v_z|)$ can be interpreted as peculiar motion of the binary is proportional
to $m_\mathrm{BH} /(m_\mathrm{BH} + m_\mathrm{LC})$.

Comparing the correlation coefficients in the delayed SN model with/without FB kick, a correlation of $(P,e)$ ($0.79$ with FB kick model, $-0.70$ without FB kick) are significant and have the opposite trend. 
Thus, we expect that we would give a constraint on the strength of natal kicks by checking the correlations obtained from the observed BH-LC samples. 

\subsection{the delayed SN model/FB kick/$\alpha=10$}
\begin{figure}[h]
	\centering
	\plotone{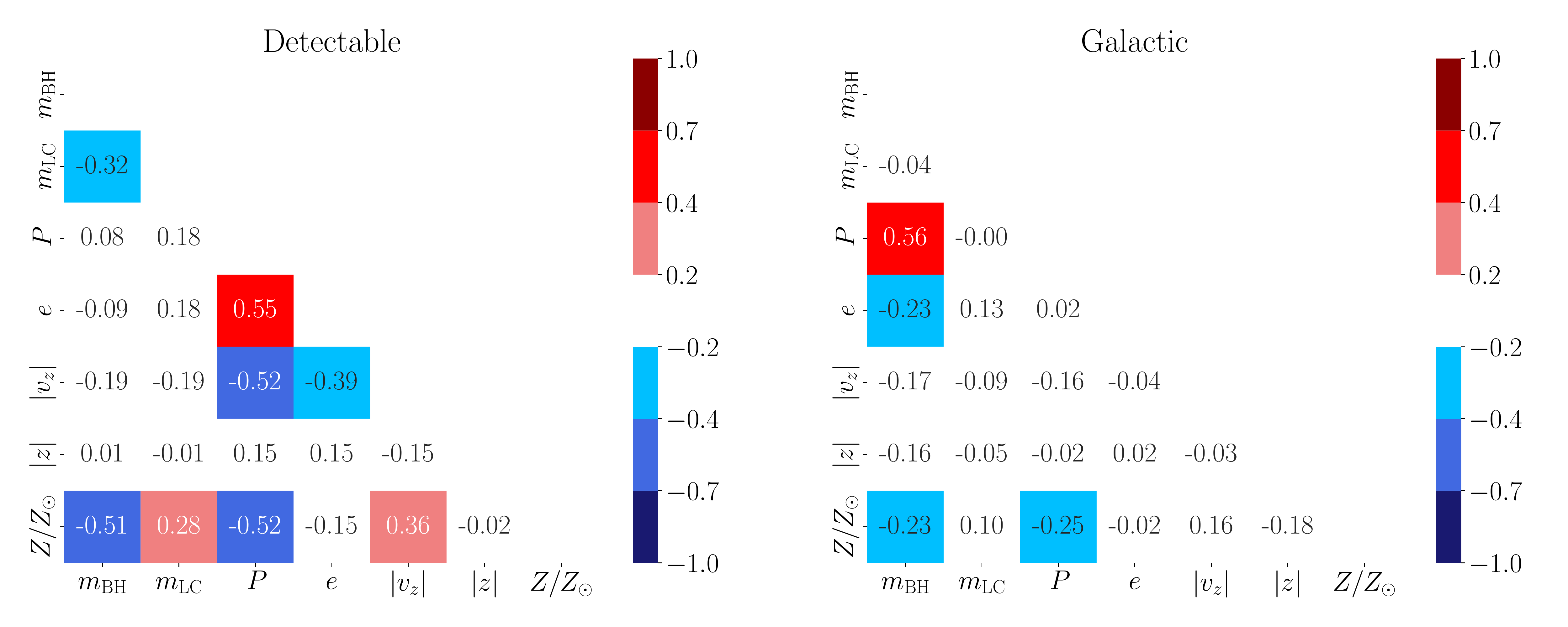}
    \caption{Same as Figure~\ref{corr_defba1}, but with a different choice of $\alpha$, $\alpha=10$.}  \label{corr_defba10}
\end{figure}

In the delayed SN model with FB kick and $\alpha=10$, 
\begin{enumerate}
    \item a positive correlation of $(P, e)$,
    \item a negative correlation of $(m_\mathrm{BH}, Z/Z_\odot)$, $(P, |v_z|)$, and $(P, Z/Z_\odot)$
\end{enumerate}
exist in Figure~\ref{corr_defba10}.
Correlations of $(P, e)$ is still seen in the higher CE efficiency case. 
Correlations of $(P, e)$ could be a clue for the strength of natal kick and SN model, regardless of the CE efficiency. 

Though the correlations seen here are similar to those for $\alpha=1$, the correlation of $(P, Z/Z_\odot)$ and $(e, Z/Z_\odot)$ have opposite signs.
That might be because the effect of FB kick seems to be different depending on the choice of $\alpha$.
In the case of $\alpha=1$, the important effect is that the FB kick, which accompanies the formation of light BHs, usually expands the orbits of binaries (see Section \ref{subsec:delay_FB_alpha1}). For a higher metallicity $Z$, the larger wind mass-loss leads to a significant reduction of the progenitor's envelope from its birth. This results in a smaller ejecta mass during the SN event, and as the kick velocity is inversely proportional to the ejecta mass the kick would be stronger. This generates the positive correlation of $(P, Z/Z_\odot)$.

On the other hand, in the case of $\alpha=10$, most of the binaries that undergo such formation histories will be disrupted during BH formation and will not contribute to the surviving population.
This is because if they experience the CE phase, their orbits will be wider for higher $\alpha$, hence are easier to disrupt. 
What survives after BH formation are binaries that barely survive the CE phase, evolving to short period BH binaries. In this case a higher metallicity, which leads to a lower envelope mass and easier envelope ejection, favors the formation of such short-period binaries and generates the negative correlation. 
 Some of their LCs finally evolve to helium-stars via Roche-lobe overflow. Though their orbits are narrow, they are relatively easier to detect due to the bright LCs. 
These binaries are found to significantly contribute to the trend of $(P, Z/Z_\odot)$ among the detectable BH binaries. 

That also alters the correlation of $(|v_z|, Z/Z_\odot)$. Combining it with the correlation of $(P, Z/Z_\odot)$, the negative correlation of $(P, |v_z|)$ could be understood.

The correlation of $(m_\mathrm{BH}, Z/Z_\odot)$ simply reflects the fact that heavy BHs are formed in low metallicities.

Thus, correlations of $(P, Z/Z_\odot)$ ($0.42$ with $\alpha = 1$, $-0.52$ with $\alpha = 10$) and $(e, Z/Z_\odot)$ ($0.71$ with $\alpha = 1$, $-0.15$ with $\alpha = 10$) could be a clue for $\alpha$.

\subsection{the rapid SN model/FB kick/$\alpha=10$}
\begin{figure}[h]
	\centering
	\plotone{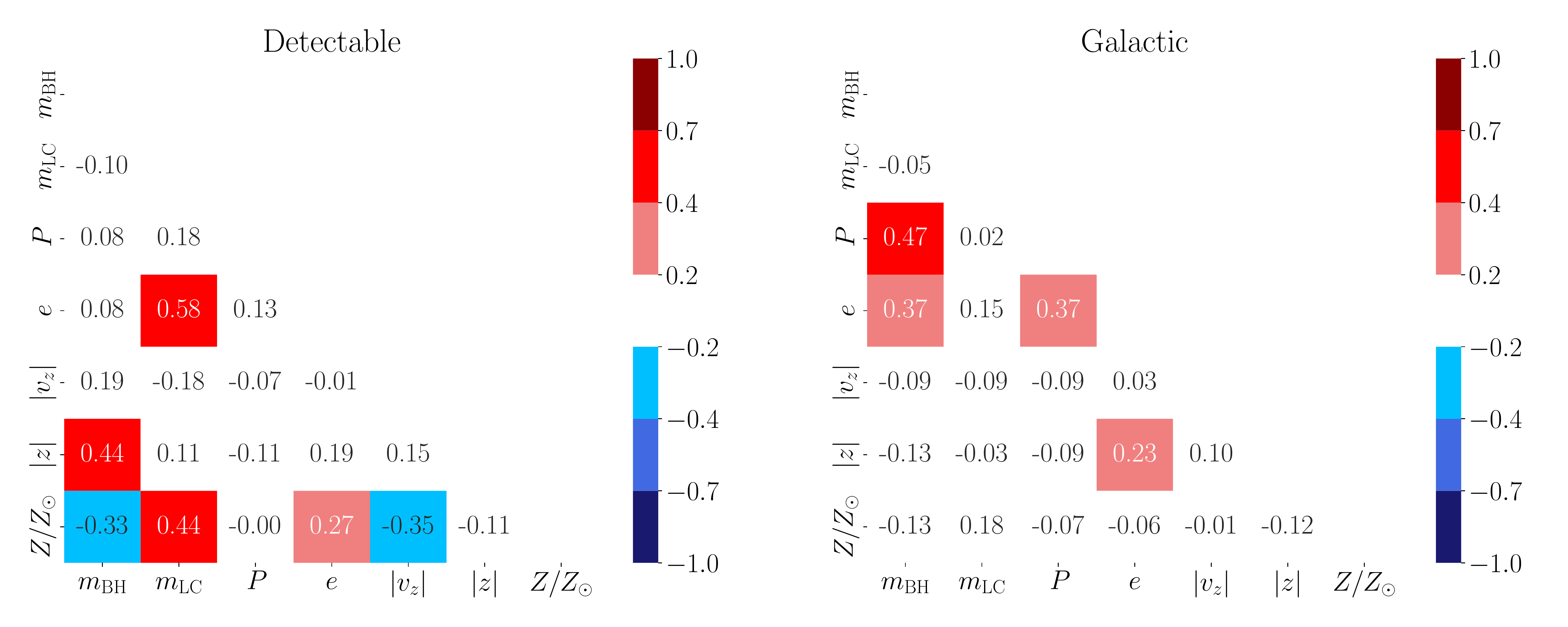}
    \caption{Same as Figure~\ref{corr_rafba1},but with a different choice of $\alpha$, $\alpha=10$.}  \label{corr_rafba10}
\end{figure}

With the high CE efficiency in the rapid SN model, we see
\begin{enumerate}
    \item positive correlations of $(m_\mathrm{BH}, |z|)$ and $(m_\mathrm{LC}, Z/Z_\odot)$, and $(m_\mathrm{LC}, e)$,
\end{enumerate}
in Figure~\ref{corr_rafba10}.
All the significant correlations follow or enhance the trend in $\alpha = 1$ case. 
The trend of $(m_\mathrm{LC}, e)$ is somehow highlighted. That might be because BH binaries with light LCs can survive the CE phase thanks to the higher $\alpha$.

As correlations of $(m_\mathrm{BH}, |z|)$ retains the same trend in $\alpha = 1$ case regardless of a choice of SN model, they would be useful to constraint SN model even if the CE efficiency is high. 

\subsection{the delayed SN model/no kick/$\alpha=10$}
\begin{figure}[h]
	\centering
	\plotone{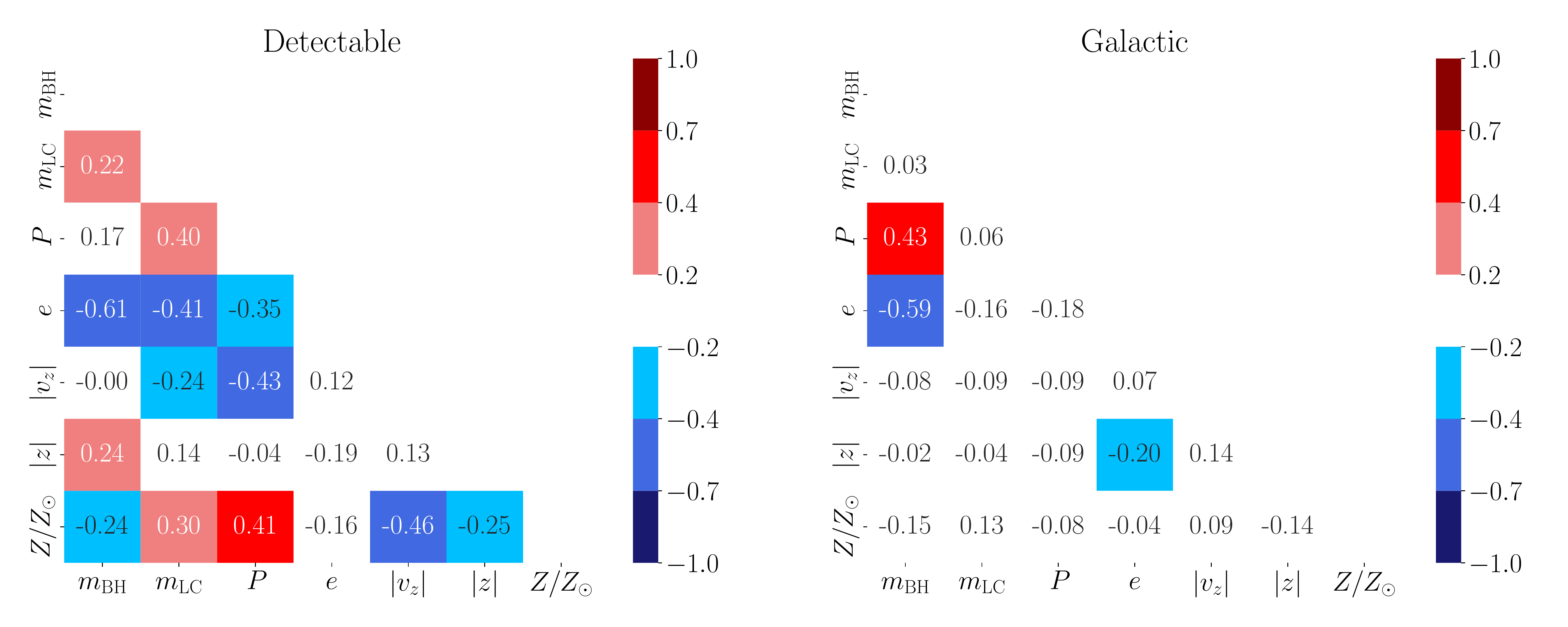}
    \caption{Same as Figure~\ref{corr_denoa1},but with a different choice of $\alpha$, $\alpha=10$.}  \label{corr_denoa10}
\end{figure}

Finally, in the delayed SN model without FB kick and $\alpha=10$, 
\begin{itemize}
    \item a positive correlation of $(P, Z/Z_\odot)$,
    \item negative correlations of $(m_\mathrm{BH}, e)$, $(m_\mathrm{LC}, e)$, $(P, |v_z|)$, $(|v_z|, Z/Z_\odot)$
\end{itemize}
are seen in Figure~\ref{corr_denoa10}. They follow the trend seen in $\alpha=1$ case. 
The correlations $(P, e)$ are somewhat blurred compared to the $\alpha = 1$ case, but still exist.
The trend of $(P, e)$ could be explained as follows. 
Due to the high CE efficiency, lighter BH binaries can survive the CE phase and their final orbits can be as wide as heavy BH binaries seen in $\alpha = 1$. That blurs the correlation of $(m_\mathrm{BH}, P)$, resulting in blurring the correlation of $(P, e)$. 
Also, the contribution of light BH binaries with larger orbital periods and a weak correlation of $(m_\mathrm{BH}, Z/Z_\odot)$ might explain the trend of $(P, Z/Z_\odot)$, which has the opposite sign seen in the lower $\alpha$ case.

Nonetheless, correlations of $(P,e)$ would give us a clue for the strength of FB kick in the high CE efficiency case. 
In addition, the correlation of $(P, Z/Z_\odot)$ may give us a clue on $\alpha$, as it has the opposite sign depending on the choice of $\alpha$.

\begin{figure}[h]
	\centering
	\plotone{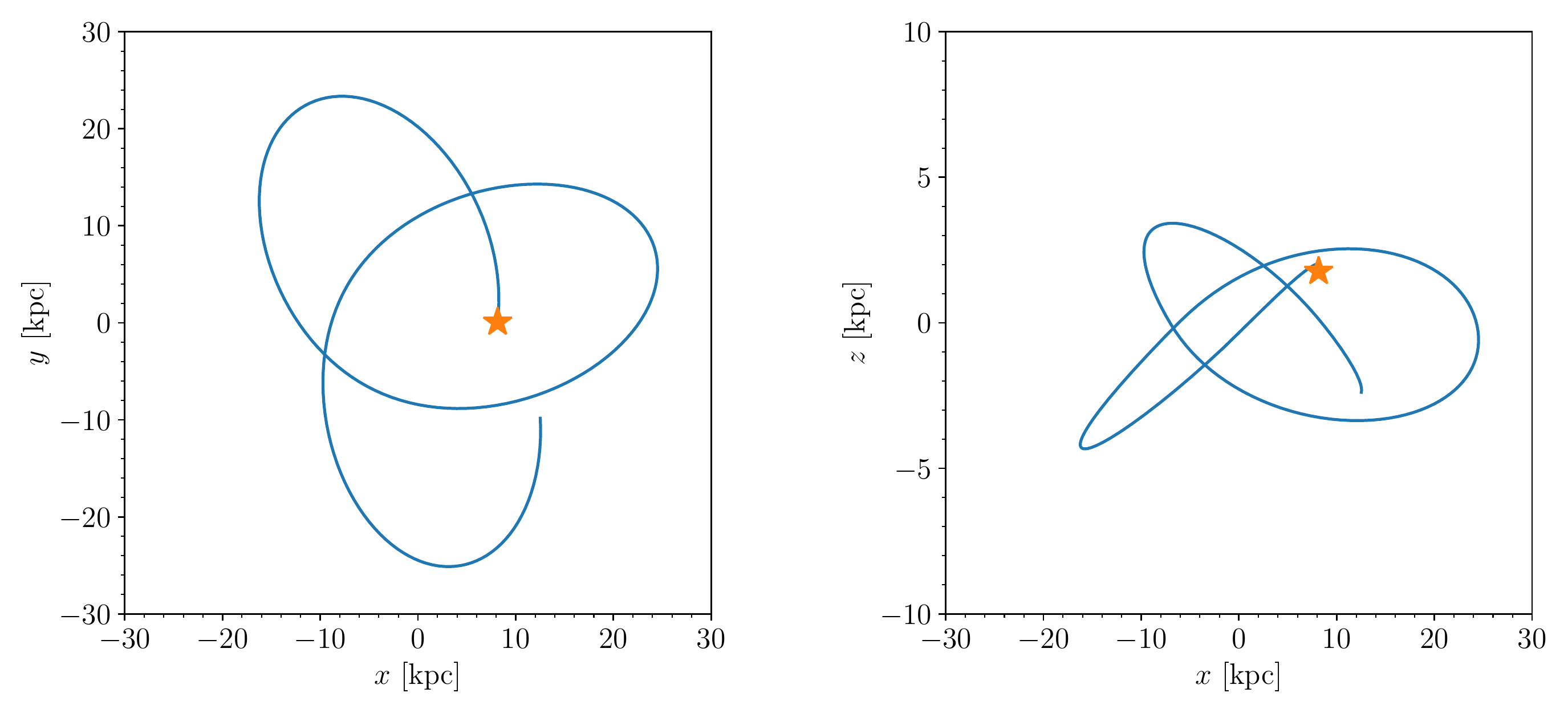}
    \caption{The same as Figure \ref{gal_path_circular}, but for a binary traveling the MW with an eccentric orbit of $e_\mathrm{gal} \sim 0.5$ (see main text for the definition of $e_\mathrm{gal}$).}  \label{gal_path_eccentric}
\end{figure}

In summary, correlations of 
\begin{itemize}
    \item $(m_\mathrm{BH}, |z|)$ ($-0.27$ in the delayed SN model, $0.42$ in the rapid SN model with $\alpha = 1$ and FB kick)
\end{itemize}
have the opposite signs by the choice of SN model. Considering mass gap BHs can be detected only in the delayed SN model, the distribution of $(m_\mathrm{BH}, |z|)$ might provide an important clue to constrain the SN model. 

Correlations of 
\begin{enumerate}
    \item $(P,e)$ ($0.79$ and $-0.70$ for delayed SN model with and without FB kick for $\alpha =1$ case respectively),
\end{enumerate}
have the opposite signs depending on the existence of FB kick. Thus, these correlations would give a constraint on the strength of natal kick. 
All of the trends summarized above would be preserved even if the CE efficiency is high.

Also, we found some correlations might give us a clue for $\alpha$ in the delayed SN model.
In the delayed SN model with FB kick, 
\begin{itemize}
    \item correlations of $(P, Z/Z_\odot)$ ($0.42$ with $\alpha = 1$, $-0.52$ with $\alpha = 10$),
    \item $(e, Z/Z_\odot)$ ($0.71$ with $\alpha = 1$, $-0.15$ with $\alpha = 10$)
\end{itemize}
have the opposite signs by a choice of $\alpha$. Correlations of $(P, Z/Z_\odot)$ would be a clue in the delayed SN model without FB kick, as well.

Finally, we investigated how eccentric are the motions of BH-LC binaries in the Galactic potential. 
A characteristic quantity we defined is ``galactic eccentricity'' $e_\mathrm{gal}$. It is defined by the maximum and the minimum radius $r_\mathrm{max}, r_\mathrm{min}$ at which each binary have reached during its lifetime, $e_\mathrm{gal} \equiv (r_\mathrm{max} - r_\mathrm{min})/(r_\mathrm{max} + r_\mathrm{min})$. 
For all the SN/kick models with $\alpha=1$, almost all ($\gtrsim 99$~\%) of the binaries have almost circular ($e_\mathrm{gal} < 0.1$) motion in the Galactic potential like shown in Figure~\ref{gal_path_circular}. In the SN models with FB kick, $0.5$~\% (delayed) and $0.1$~\% (rapid) binaries have eccentric orbits with $e_\mathrm{gal} > 0.5$. An example of the Galactic path for one of the binaries is shown in Figure~\ref{gal_path_eccentric}.

\section{Comparison with the Confirmed BH Binary and the BH Candidates with \textit{Gaia}} 
\label{sec:discussion}
In this section, we compare our results with BH candidates reported in \textit{Gaia} DR3 
by selecting some candidates among them and discussing how they can be formed from isolated field binaries. 

Including \textit{Gaia} BH 1, we select the candidates from \cite{Andrewsetal2022, Shahafetal2022, Tanikawaetal2022} with the upper limit of compact object mass larger than $3 M_\odot$ from \cite{Andrewsetal2022, Shahafetal2022, Tanikawaetal2022} since \textsf{BSE} considers compact objects heavier than $3 M_\odot$ as BHs. 
These BH candidates can be roughly divided into two types in terms of component mass and orbital period: one is $\lesssim 4 M_\odot$ BH and $\lesssim 1.5 M_\odot$ LC binaries with long orbital periods ($P \sim 1.5 - 4$~years) and non-zero eccentricities (type 1) and the other is $\gtrsim 9 M_\odot$ BH and $\lesssim 1.2 M_\odot$ LC binaries with short orbital periods ($P \lesssim 1$~year) and non-zero eccentricities (type 2). The latter type includes \textit{Gaia} BH 1. We summarize the BH candidates and \textit{Gaia} BH 1 in Table~\ref{cand_info}. We note that the LC mass of the candidate reported in \cite{Tanikawaetal2022} is not estimated, so we do not categorize it as either type.

\begin{table}[h]
	\centering
\begin{tabular}{c|ccccc} \hline
\textit{Gaia} ID & BH mass $[M_\odot]$ & LC mass $[M_\odot]$ & $P$~[days] & $e$ & type \\ \hline 
$4314242838679237120^{*1}$ & $2.25^{+1.87}_{-0.84}$ & $0.63 - 1.00$ & $1146 \pm 382$ & $0.70 \pm 0.09$ & $1$\\ 
$5593444799901901696^{*1}$ & $2.57^{+0.86}_{-0.69}$ & $1.27 \pm 0.2$ & $1039 \pm 292$ & $0.44 \pm 0.14$ & $1$\\
$6328149636482597888^{*1}$ & $2.71^{+1.50}_{-0.36}$ & $1.21 \pm 0.2$ & $736 \pm 23$ & $0.14 \pm 0.07$ & $1$\\
$6281177228434199296^{*2}$ & $11.9 \pm 1.5$ & $1.0$ & $153.95 \pm 0.36$ & $0.180 \pm 0.042$ & $2$\\
$3509370326763016704^{*2}$ & $3.69 \pm 0.24$ & $0.7$ & $109.392 \pm 0.065$ & $0.237 \pm 0.016$ & $1$\\
$6802561484797464832^{*2}$ & $3.08 \pm 0.84$ & $1.2$ & $574.8 \pm 6.2$ & $0.830\pm 0.071$ & $1$\\
$3263804373319076480^{*2}$ & $2.75 \pm 0.50$ & $1.0$ & $510.7 \pm 4.7$ & $0.278 \pm 0.023$ & $1$\\
$6601396177408279040^{*2}$ & $2.57 \pm 0.50$ & $1.0$ & $533.5 \pm 2.0$ & $0.791 \pm 0.043$ & $1$\\
$4373465352415301632^{*3}$ & $9.78 \pm 0.18$ & $0.93 \pm 0.05$ & $185.63 \pm 0.05$ & $0.454 \pm 0.005$ & $2$ \\
$5870569352746779008^{*4}$ & $> 5.25$ & & $1352.25 \pm 45.50$ & $0.5324 \pm 0.0095$ & \\ \hline
\end{tabular}
	\caption{Information of the BH candidates reported in \textit{Gaia} DR 3 whose BH mass exceeds $3 M_\odot$ at the upper limit and \textit{Gaia} BH 1 \citep{ElBadryetal2023}. Binary parameters of the candidates are based on the \textit{Gaia} DR3 database. We cited parameters estimated in \cite{ElBadryetal2023} for the information of \textit{Gaia} BH 1. \\ $*1$: reported in \cite{Andrewsetal2022}, $*2$: reported in \cite{Shahafetal2022}, $*3$: reported in \cite{ElBadryetal2023}, and $*4$: \cite{Tanikawaetal2022}.}
 \label{cand_info}
\end{table}

\begin{figure}[h]
	\centering
	\plotone{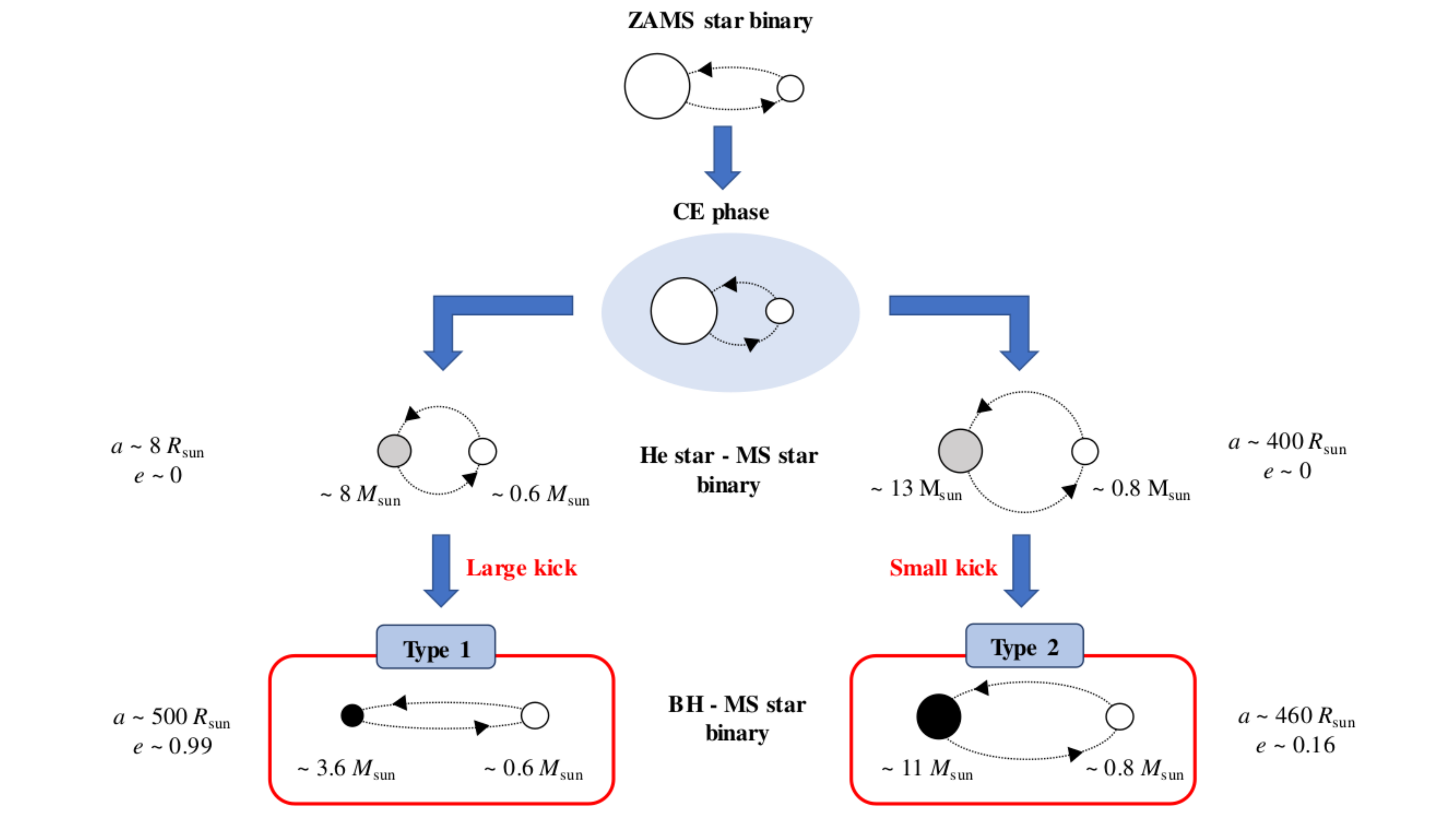}
    \caption{An example of evolutionary paths of type 1-like BH binary (\textit{e.g}~$\lesssim 4 M_\odot$ BH and $\lesssim 1.5 M_\odot$ LC binaries with long orbital periods ($P \sim 1.5 - 4$~years) and non-zero eccentricities) and type 2-like BH binary (\textit{e.g.}~$\gtrsim 9 M_\odot$ BH and $\lesssim 1.2 M_\odot$ LC binaries with short orbital periods ($P \lesssim 1$~year) and non-zero eccentricities). Both types of binaries experience the CE phase and finally form different range of BH mass, which makes a difference in terms of orbital separations and eccentricities depending on the strength of mass loss kick. }  \label{evo_path}
\end{figure}

We found that the delayed SN model with no natal kick and $\alpha=10$ stably forms both types of BH binaries. Based on our simulation, $\sim 3 \times 10^4$ type 1-like binaries and $\sim 900$ type 2-like binary are expected to exist in the MW. Figure \ref{evo_path} shows examples of evolutionary path for both types of BH binaries in the delayed SN model with no kick and $\alpha = 10$. 
The evolutionary path for both types of binaries is almost the same: they experience the CE phase before forming BHs. The difference is that BH mass of type 1-like binaries is lighter. Thus, their mass loss kick (\textit{i.e.}~the Blaauw kick) is larger than that of type 2-like ones, which makes orbits of type 1-like binaries wider and more eccentric.
The delayed SN models with FB kick and $\alpha=10$ may also form both types of binaries. While type-1 like binaries are stably formed, type-2 like binaries were sometimes born if the strength of natal kick is relatively small such as tens of $\SI{}{km. s^{-1}}$ to $\SI{135}{km. s^{-1}}$. The existence of natal kick can make the orbits of type 2-like binaries more eccentric ($e \sim 0.3 - 0.6$) and narrower ($P \sim 200$~days), more similar to \textit{Gaia} BH 1.

However, the other models, \textit{i.e.}~the rapid SN model regardless of the CE efficiency or the delayed SN model with the low CE efficiency, cannot form both types of binaries.
Light BH binaries with long orbital periods cannot be formed in the rapid SN model. Some BHs as light as $\lesssim 4 M_\odot$ are formed in the rapid SN model via accretion-induced collapse, but their orbital periods are shorter than $1$~year. 
Thus, if we confirm that the candidates of type 1-like binaries possess BHs, SN models producing mass gap BHs like the delayed SN model is favored.
In the delayed SN model with low CE efficiency, if one attempts to form Type 2-like binaries with heavier BHs, their final orbital periods become $\sim 10$~days, much shorter than observed. 

\section{Conclusion} 
\label{sec:conclusion}
We investigated correlations between binary parameters (BH mass, LC mass, orbital periods, and eccentricities), spatial parameters (velocities perpendicular to the Galactic plane, and the heights from the Galactic plane), and metallicity of BH-LC binaries detectable with \textit{Gaia}.
In this work, we explore the effects of choices of binary evolution models, SN model, BH natal kick, and CE efficiency, which are of great uncertainties on the number of detectable BH binaries as \cite{Shikauchietal2022} have estimated the number may vary by 40 times by choices of binary evolution models.
By sampling initial spatial conditions, metallicity and lookback time distributions based on \cite{Waggetal2021}, then simulating binary evolution with \textsf{BSE} and the orbit of the binary under the Galactic potential, we obtained the BH-LC binary population in the MW.

We conclude that most of the correlation coefficients among the detectable binaries have the opposite sign and/or are enhanced by the detection criteria as correlation coefficients among the Galactic population show almost no correlations. 
Nevertheless, we indicated some correlations might probe the SN model and the strength of natal kick regardless of the CE efficiency.
Correlations of $(m_\mathrm{BH}, |z|)$ would be a clue for the SN model if a strong natal kick like FB kick exists. In the delayed SN model light BHs ($m_\mathrm{BH} \lesssim 4 M_\odot$) are formed, and binaries possessing such BHs go farther from the Galactic plane due to strong kick, resulting in a negative correlation. On the other hand, in the rapid SN model, light BHs are rarely formed and natal kick is not so strong, thus BH binaries do not leave far away from the Galactic plane. The detection criteria simply emphasizes heavier BH binaries can be detected at farther distances. 

The signs of correlations of $(P,e)$ vary depending on the existence of FB kick, which would be useful to constraint the strength of natal kick. 
With FB kick, light BH binaries suffer from a strong kick, which makes their orbits more eccentric and wider. 
On the other hand, due to the absence of FB kick, light BH binaries can remain tighter than those in the same SN model with FB kick, resulting in a positive correlation of $(m_\mathrm{BH}, P)$ and the opposite correlation of $(P, e)$.

Moreover, in the delayed SN model correlations of $(P, Z/Z_\odot)$ and/or $(e, Z/Z_\odot)$ would be an indicator of the CE efficiency $\alpha$.
The reason we see such a trend is different by the choice of the natal kick model. 
For the model with FB kick, the kick expands light BH binaries' orbits in lower $\alpha$ case. On the other hand, it rather disrupts most of such binaries at BH formation in higher $\alpha$.
For the model without FB kick, light BH binaries simply can have as wide orbits as heavy BH binaries in higher $\alpha$ case. 

We note that BSE takes into account stellar evolution in a rather simple formalism, as is the case for other rapid binary population synthesis calculations. For example, uncertainties in rotation or convection can affect the mass of the helium and carbon-oxygen cores. These uncertainties can affect the remnant mass, and possibly alter the distribution of BH mass from what is considered here.

Using BH-LC samples we employed here, we also investigated the possibility of forming binaries like the BH candidates reported in \textit{Gaia} DR3 \citep{Andrewsetal2022, Shahafetal2022} and \textit{Gaia} BH 1 \citep{ElBadryetal2023} in each SN/kick model with a choice of $\alpha$ used in this work.
We divided all the candidates and \textit{Gaia} BH 1 into two groups, type 1 and 2 (see Table~\ref{cand_info}), in terms of component masses and orbital periods. 
We revealed that only the delayed SN model with the high CE efficiency can form both types of binaries in an isolated field.
Both types of binaries are formed via the CE phase. 
If the CE efficiency is as low as unity, type 2-like binaries can not have as large orbital separations as the observed ones.
Especially, the rapid SN model cannot form type 1-like binaries since such light BHs are formed via accretion-induced collapse, which requires shorter orbital separations than seen in type 1-like binaries. 
We also expect the SN model producing light BHs of masses $\lesssim 4 M_\odot$ would be favored if BH candidates categorized as type 1 binaries are confirmed as genuine BH binaries. 

\begin{figure}[h]
	\centering
	\plotone{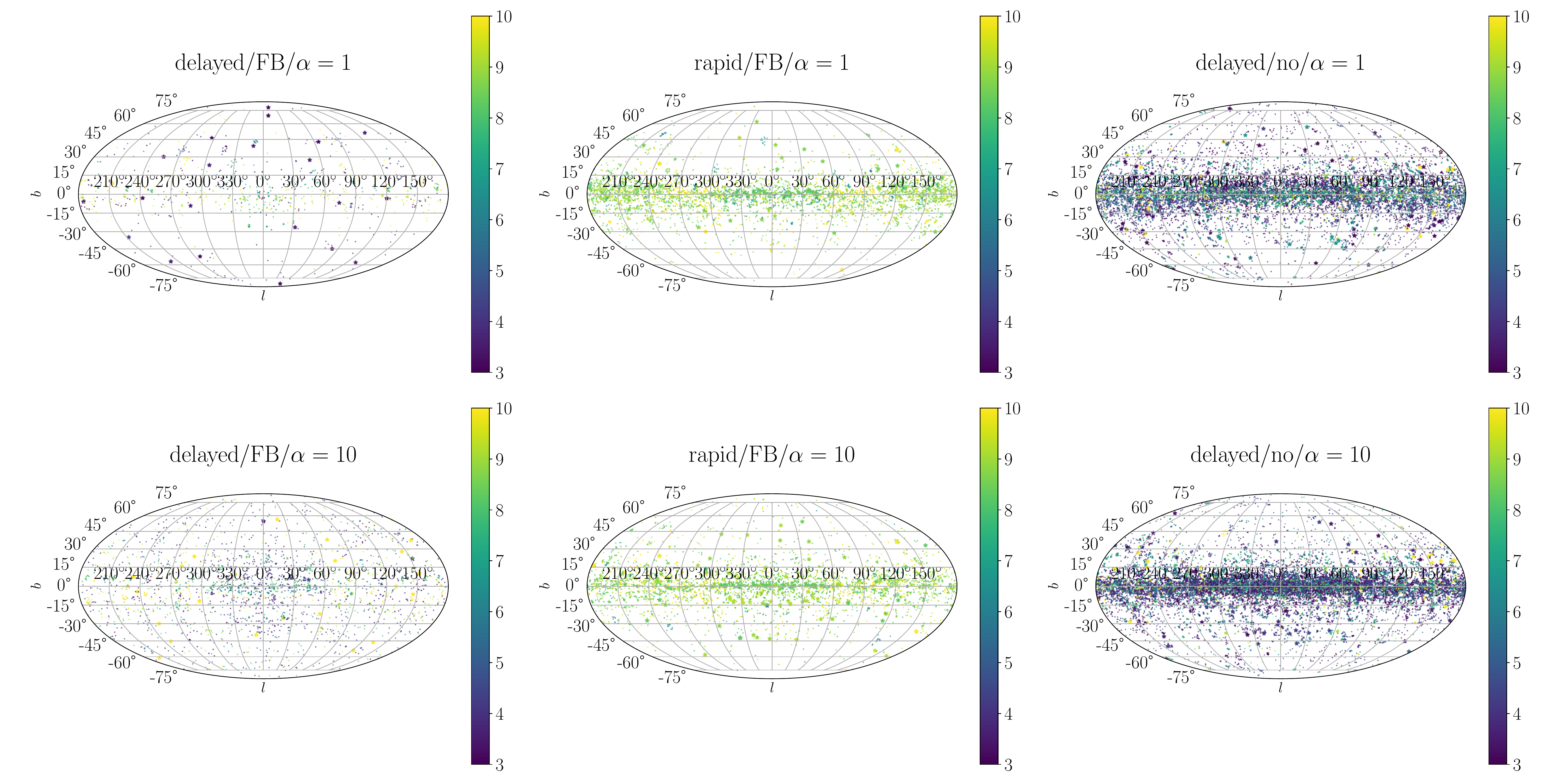}
    \caption{Spatial distributions of the detectable BH-LC binaries obtained from 10 different realizations with different choices of the SN/kick models and values of the CE efficiency $\alpha$.
    Maps are shown in the Galactic coordinate. Each star marker shows each binary, whose size is proportional to the weighting factor. Colors of each marker correspond to BH mass.}  \label{prospect_2}
\end{figure}

\begin{figure}[h]
	\centering
	\plotone{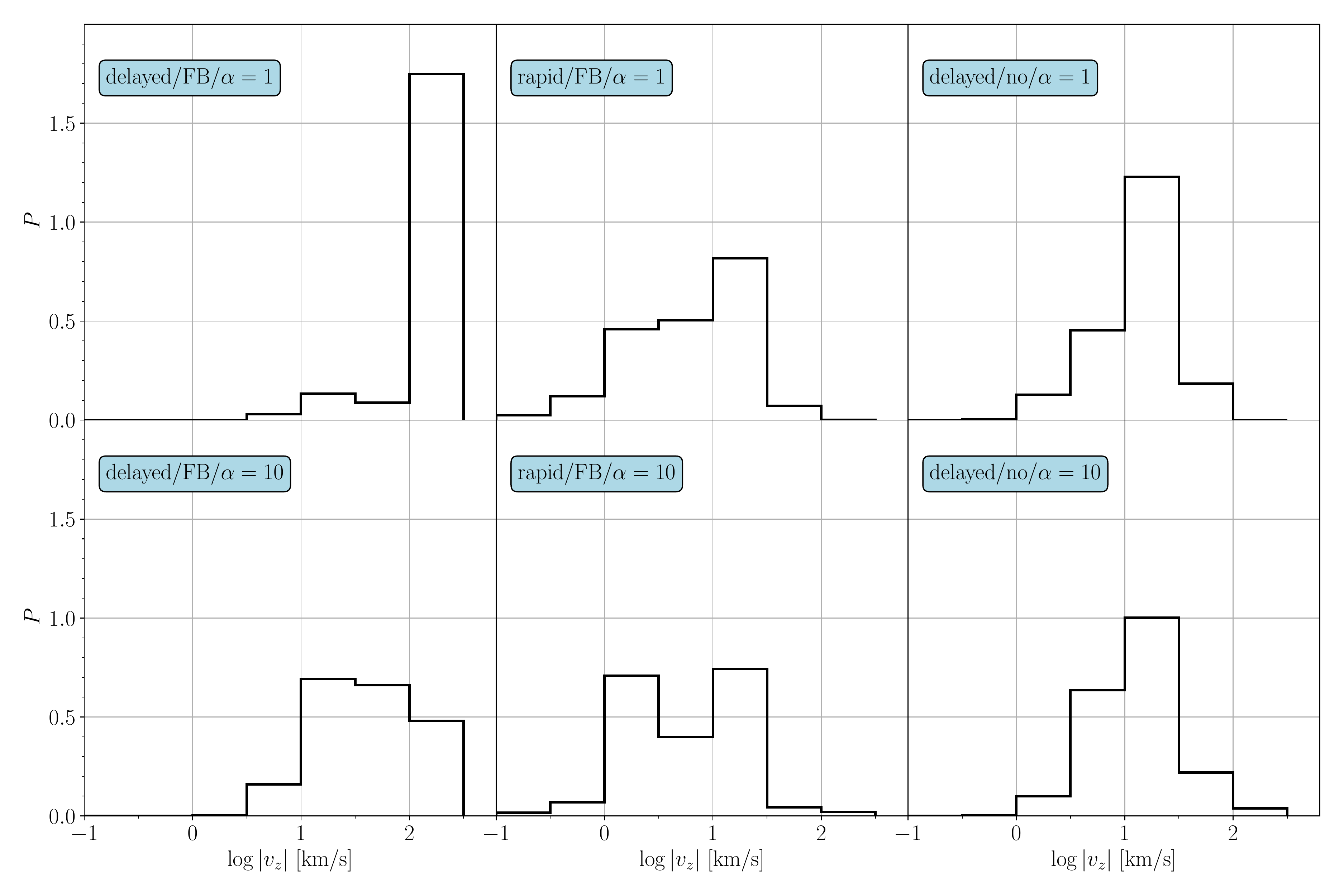}
    \caption{Probability density function of $|v_z|$ of the detectable BH-LC binaries with different choices of the SN/kick models and values of the CE efficiency $\alpha$. }  \label{vz_hist}
\end{figure}

As more candidates are identified as genuine BH binaries, spatial distributions of BH-LC binaries in the Galactic coordinate will be obtained as shown in Figure~\ref{prospect_2}. 
Each point in the figure depicts the detectable BH-LC binaries obtained from all the realizations, weighted by the weighting factor and colored by BH mass. 
In the delayed SN model, light BHs ($m_\mathrm{BH} \lesssim 5 M_\odot$) would be detectable at a high longitude such as $|b| > 45^{\circ}$. 
Also, we expect $|v_z|$ distributions might tell us the strength of natal kick. 
Figure~\ref{vz_hist} shows a probability density function of $\log |v_z|$ of the detectable BH-LC binaries. 
If a strong natal kick model such as FB kick is favored, $\gtrsim 50$~\% of the BH-LC binaries would have a large $|v_z|$ such as $\sim \SI{30}{km. s^{-1}}$.
If SN model which does not produce lower mass gap BHs is favored, the detected BH-LC binaries are less likely to have such a large $|v_z|$. 

\section*{Acknowledgement}
We thank the anonymous referee for useful comments and Kareem El-Badry for useful discussions.

M.S. is supported by Research Fellowships of Japan
Society for the Promotion of Science for Young Scientists, by Forefront Physics and Mathematics Program to Drive Transformation (FoPM), a World-leading Innovative Graduate Study (WINGS) Program, the University of Tokyo, and by JSPS Overseas Challenge Program for Young Researchers. D.T. is supported by the Sherman Fairchild Postdoctoral Fellowship at Caltech. This research is supported by Grants-in-Aid for Scientific Research (17H06360, 19K03907, 22K03686) from the Japan Society for the Promotion of Science.

\appendix
\section{Corner Plots of Binary Parameters}
\label{app:corr}
\begin{figure}[h]
	\centering
	\plotone{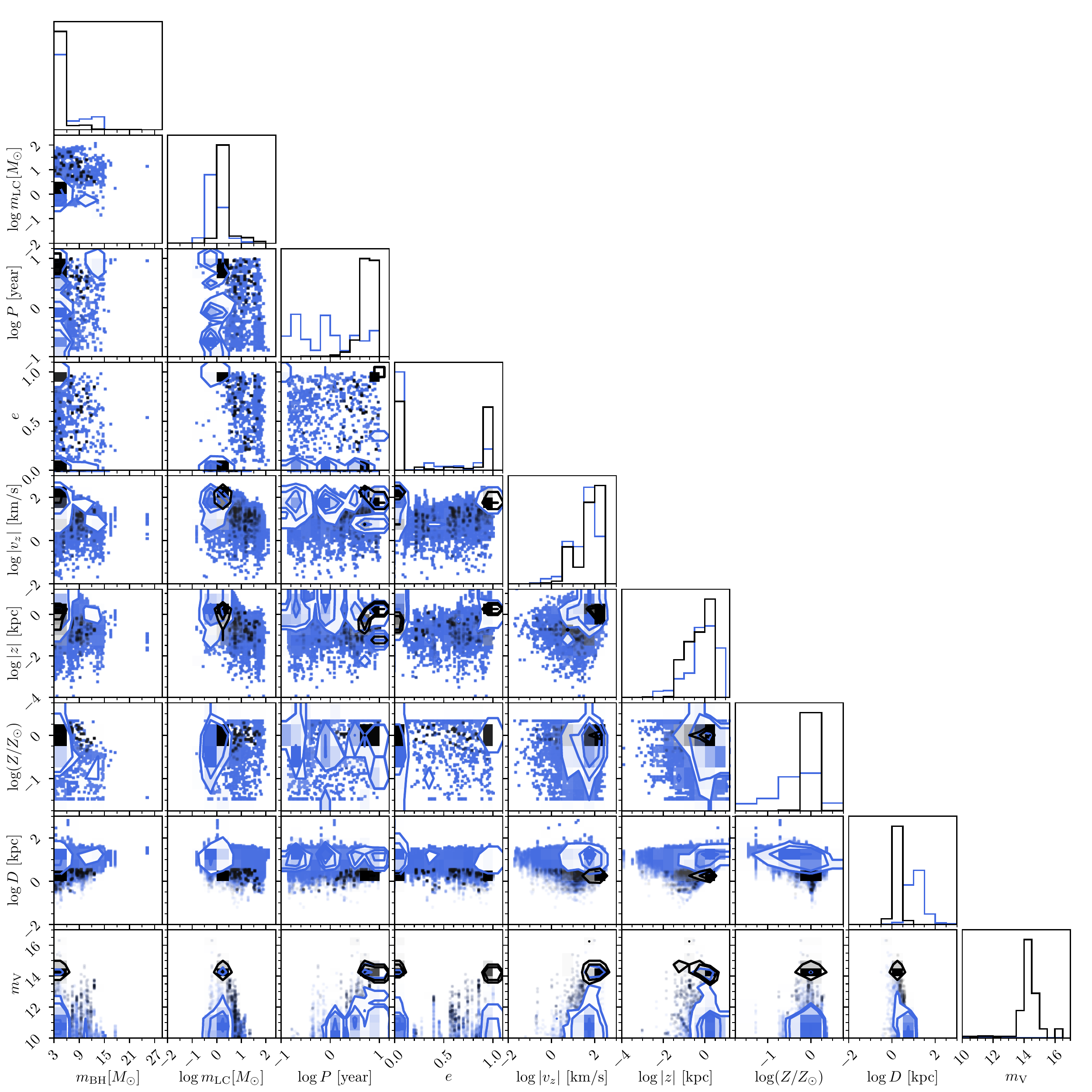}
    \caption{Corner plot of the current binary parameters (BH mass $m_\mathrm{BH}$, LC mass $m_\mathrm{LC}$, orbital periods $P$, and eccentricities $e$), the current spatial parameters (velocities in $z$-direction $|v_z|$, and the heights from the Galactic plane $|z|$), metallicity $Z/Z_\odot$, distances from the Earth to BH binaries $D$ and apparent magnitudes $m_\mathrm{V}$ with the delayed SN model including FB kick and $\alpha=1$. Histograms show the distribution of each parameter. The black lines and contours show the distributions of the detectable BH-LC binaries. The blue ones correspond to the Galactic BH-LC binaries within a period range, $P = 50$~days to $10$~years.}  \label{cp_de_fb}
\end{figure}

\begin{figure}[h]
	\centering
	\plotone{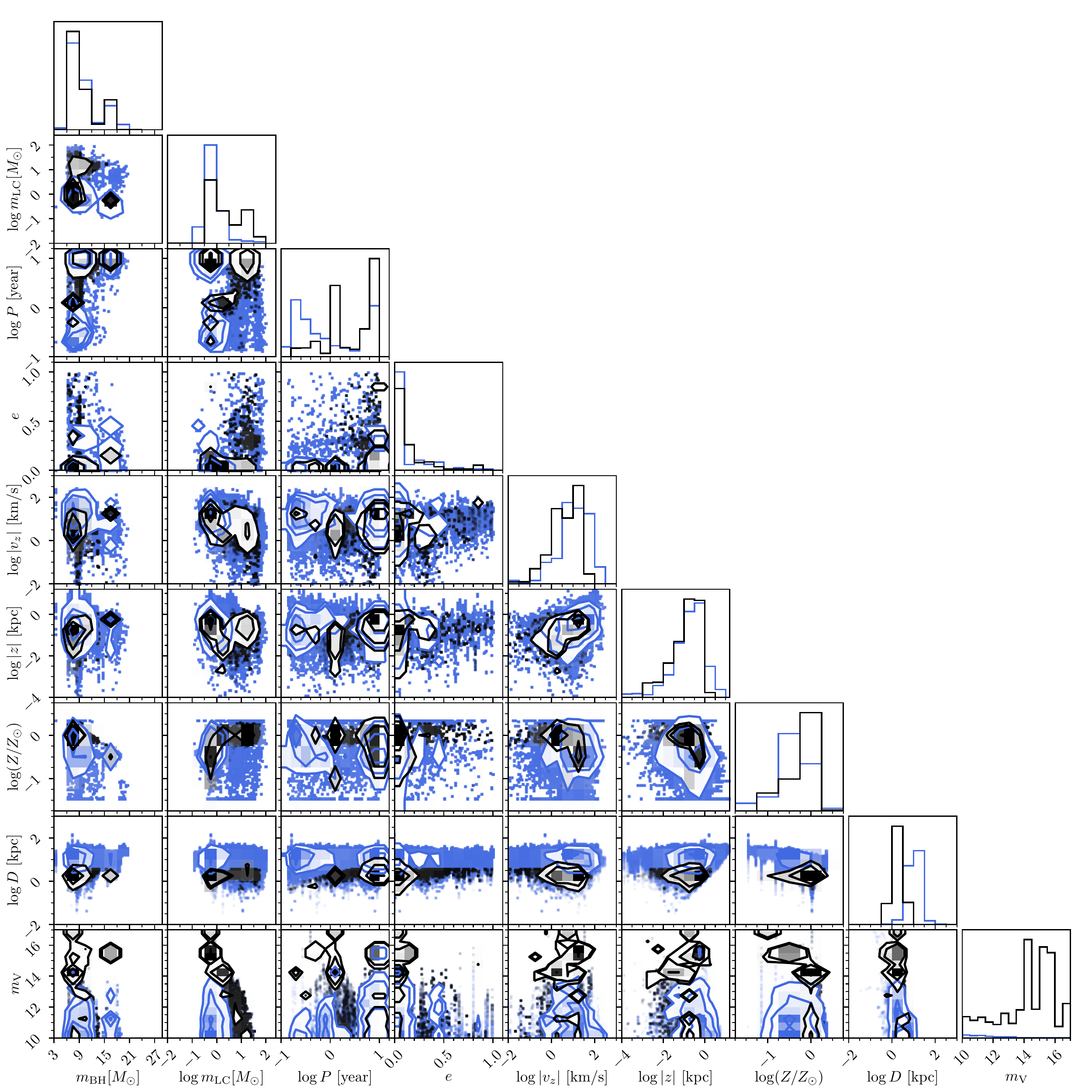}
    \caption{The same as Figure \ref{cp_de_fb} except for the SN model. Here, the rapid SN model was employed.}  \label{cp_ra_fb}
\end{figure}

\begin{figure}[h]
	\centering
	\plotone{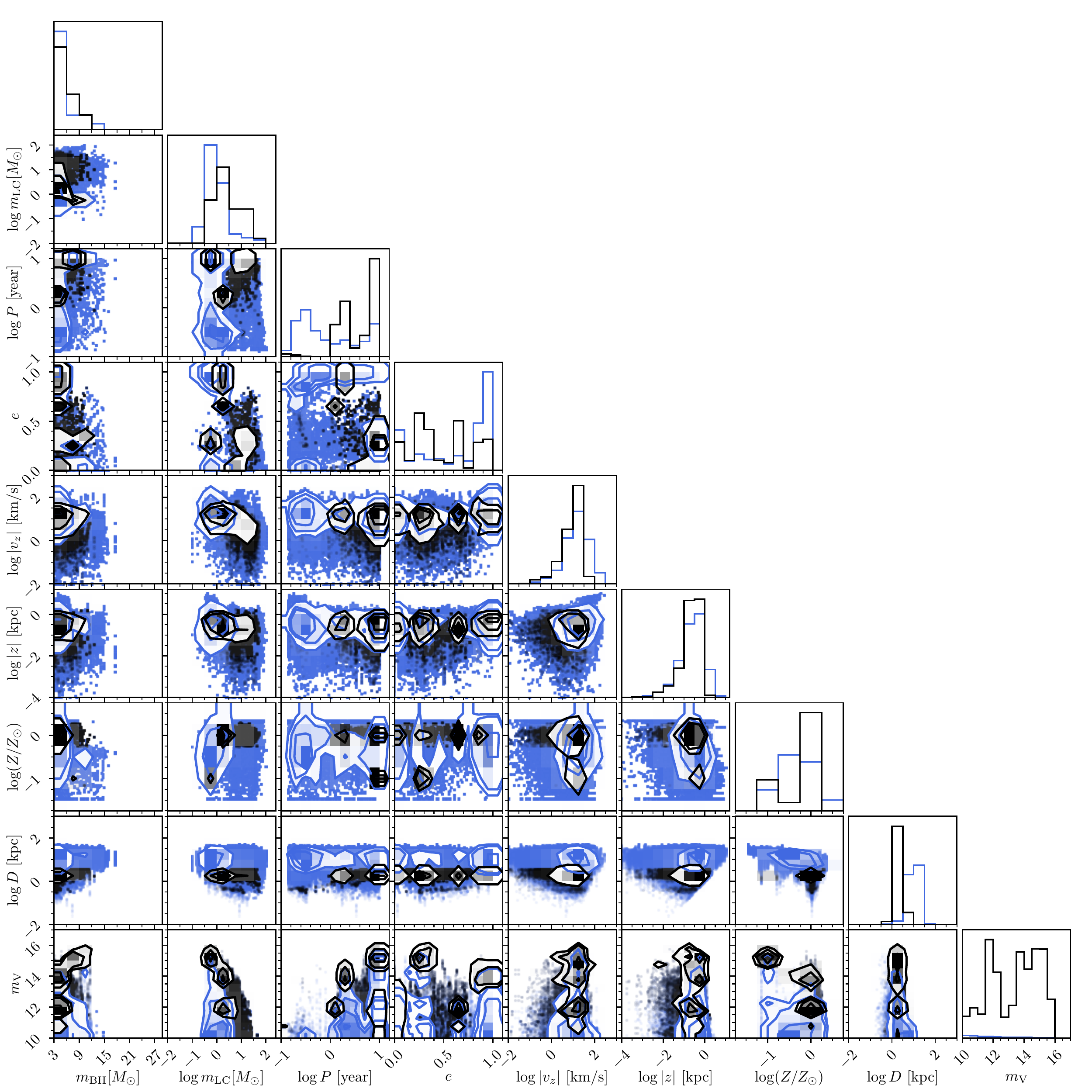}
    \caption{The same as Figure \ref{cp_de_fb} except for kick model. Natal kick was excluded.}  \label{cp_de_no}
\end{figure}

\begin{figure}[h]
	\centering
	\plotone{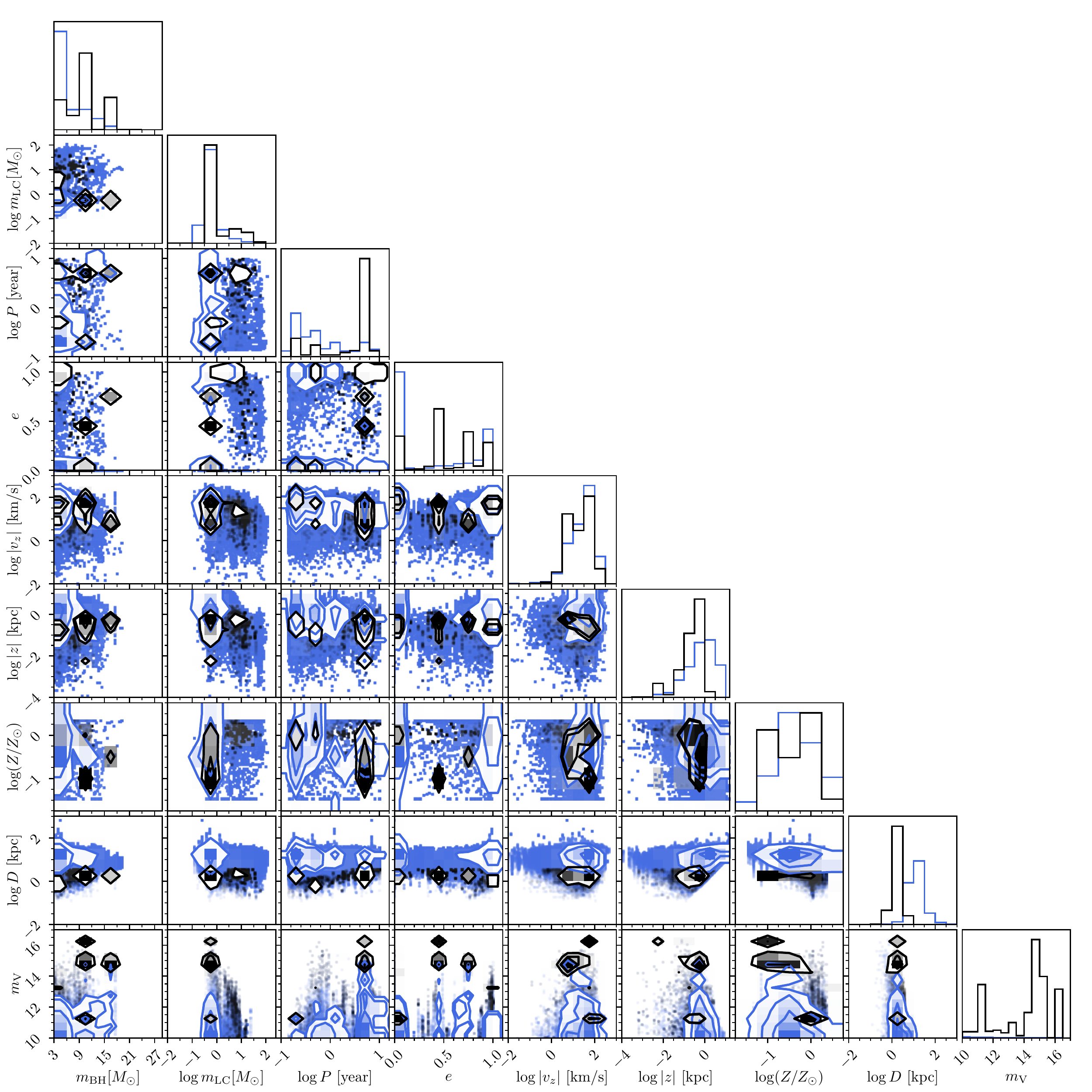}
    \caption{The same as Figure \ref{cp_de_fb} except for the CE efficiency. Here, we adopted $\alpha=10$.}  \label{cp_defba10}
\end{figure}

\begin{figure}[h]
	\centering
	\plotone{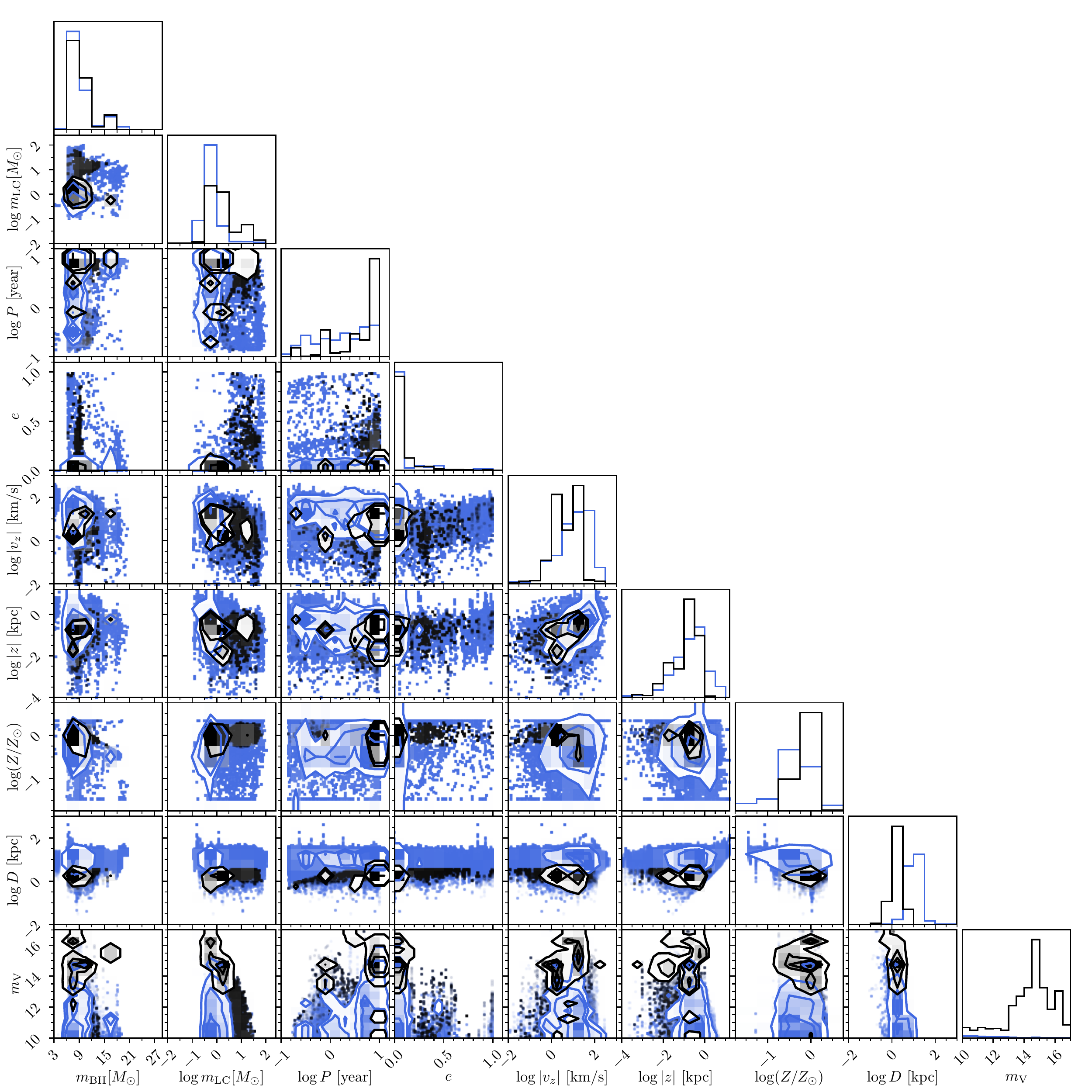}
    \caption{The same as Figure \ref{cp_ra_fb} except for the CE efficiency. Here, we adopted $\alpha=10$.}  \label{cp_rafba10}
\end{figure}

\begin{figure}[h]
	\centering
	\plotone{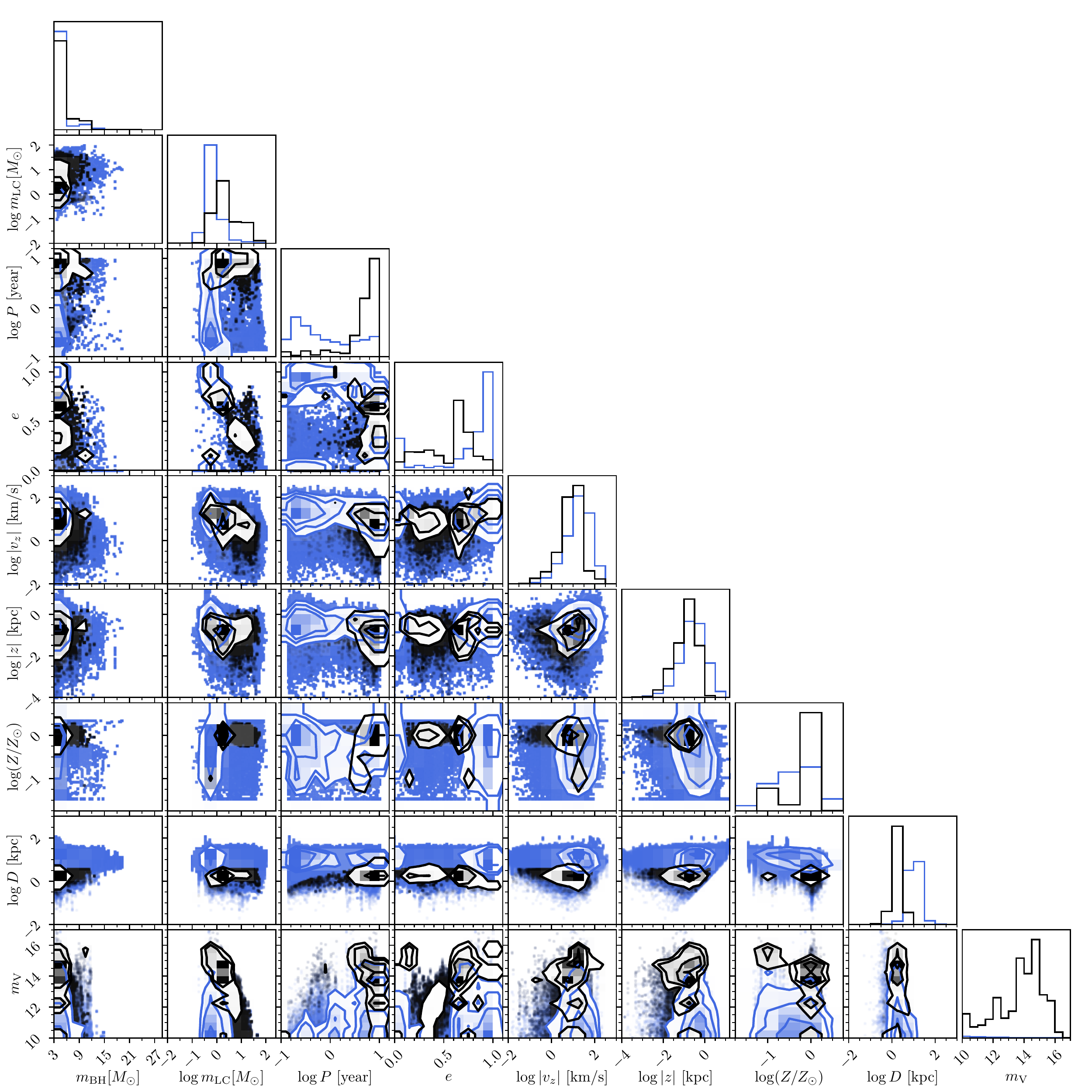}
    \caption{The same as Figure \ref{cp_de_no} except for the CE efficiency. Here, we adopted $\alpha=10$.}  \label{cp_denoa10}
\end{figure}

In this appendix, Figures~\ref{cp_de_fb} to \ref{cp_denoa10} show two-dimensional scatter plots with binary parameters (BH mass $m_\mathrm{BH}$, LC mass $m_\mathrm{LC}$, orbital periods $P$, and eccentricities $e$), spatial parameters (velocities in $z$-direction $|v_z|$, and the heights from the Galactic plane $|z|$) metallicity besides distances from the Earth to BH binaries $D$ and apparent magnitudes $m_\mathrm{V}$, and one-dimensional histograms for each choice of SN/kick models and $\alpha$ values. Note that the vertical axis in the histograms is linear.
Each point shows a BH-LC sample obtained from the 10 different realizations. 
The black point depicts the detectable BH-LC binaries. 
The blue ones are the entire Galactic BH-LC binaries with orbital period of $50$~days to $10$~years.

\bibliography{references}{}
\bibliographystyle{aasjournal}



\end{document}